\documentclass[aps,showpacs,pre,twocolumn,superscriptaddress,floatfix]{revtex4}
 \usepackage{graphicx,bm,amssymb,amsmath,dcolumn,hyperref}
 \usepackage{multirow}
 \usepackage{color}
 \usepackage{subfigure}  %added.
 \usepackage{epstopdf}
 \usepackage{bbm}
 \usepackage{graphicx}
 \usepackage{hyperref}
 \usepackage{threeparttable}
 \usepackage{amsthm}

\begin{document}
\newcommand{\be}{\begin{equation}}
\newcommand{\ee}{\end{equation}}
\newcommand{\half}{\frac{1}{2}}
\newcommand{\ith}{^{(i)}}
\newcommand{\im}{^{(i-1)}}
\newcommand{\gae}
{\,\hbox{\lower0.5ex\hbox{$\sim$}\llap{\raise0.5ex\hbox{$>$}}}\,}
\newcommand{\lae}
{\,\hbox{\lower0.5ex\hbox{$\sim$}\llap{\raise0.5ex\hbox{$<$}}}\,}

\definecolor{blue}{rgb}{0,0,1}
\definecolor{red}{rgb}{1,0,0}
\definecolor{green}{rgb}{0,1,0}
\newcommand{\blue}[1]{\textcolor{blue}{#1}}
\newcommand{\red}[1]{\textcolor{red}{#1}}
\newcommand{\green}[1]{\textcolor{green}{#1}}

\newcommand{\scrA}{{\mathcal A}}
\newcommand{\scrE}{{\mathcal E}} %added.
\newcommand{\scrF}{{\mathcal F}} %added.
\newcommand{\scrG}{{\mathcal G}}
\newcommand{\scrL}{{\mathcal L}}
\newcommand{\scrM}{{\mathcal M}} %added.
\newcommand{\scrN}{{\mathcal N}}
\newcommand{\scrS}{{\mathcal S}}
\newcommand{\scrs}{{\mathcal s}}
\newcommand{\scrP}{{\mathcal P}}
\newcommand{\scrO}{{\mathcal O}}
\newcommand{\scrR}{{\mathcal R}}
\newcommand{\scrC}{{\mathcal C}}
\newcommand{\scrV}{{\mathcal V}}
%added.
\newcommand{\dm}{d_{\rm min}}
\newcommand{\db}{d_{\rm B}}
\newcommand{\df}{d_{\rm F}}
\newcommand{\bx}{{\bf x}}
\newcommand{\by}{{\bf y}}
\title{Geometric properties of the Fortuin-Kasteleyn representation of the Ising model}
\date{\today}

\date{\today}
\author{Pengcheng Hou}
\affiliation{Hefei National Laboratory for Physical Sciences at Microscale and 
Department of Modern Physics, University of Science and Technology of China, 
Hefei, Anhui 230026, China}
\author{Sheng Fang}
\affiliation{Hefei National Laboratory for Physical Sciences at Microscale and 
Department of Modern Physics, University of Science and Technology of China, 
Hefei, Anhui 230026, China}
\author{Junfeng Wang}
\email{wangjf@hfut.edu.cn}
\affiliation{School of Electronic Science and Applied Physics, Hefei University 
of Technology, Hefei, Anhui 230009, China}
\author{Hao Hu}
\email{huhao@ahu.edu.cn}
\affiliation{School of Physics and Materials Science, Anhui University, Hefei 230601, China}
\author{Youjin Deng}
\email{yjdeng@ustc.edu.cn}
\affiliation{Hefei National Laboratory for Physical Sciences at Microscale and 
Department of Modern Physics, University of Science and Technology of China, 
Hefei, Anhui 230026, China}
\affiliation{CAS Center for Excellence and Synergetic Innovation Center in Quantum Information and Quantum Physics, University of Science and Technology of China, Hefei, Anhui 230026, China}

\begin{abstract}
	We present a Monte Carlo study of the Fortuin-Kasteleyn (FK) clusters of the Ising model
	on the square (2D) and simple-cubic (3D) lattices. The wrapping probability, a dimensionless quantity
	characterizing the topology of the FK clusters on a torus, is found to suffer from smaller finite-size
	corrections than the well-known Binder ratio, and yields a high-precision critical coupling as 
	$K_c(3\rm D)=0.221\,654\,631(8)$. 
	We then study geometric properties of the FK clusters at criticality.
	It is demonstrated that the distribution of the critical   largest-cluster
	size $C_1$ follows a single-variable function as $P(C_1,L){\rm d}C_1=\tilde P(x){\rm d}x$ with
	$x\equiv C_1/L^{\df}$ ($L$ is the linear size), and that the fractal dimension $\df$ is identical 
	to the magnetic exponent. An interesting bimodal feature is observed in distribution $\tilde P(x)$ in 3D,
	and attributed to the different approaching behaviors for $K \to K_c+0^\pm$. 
	For a critical FK configuration, the cluster number per site $n(s,L)$ of size $s$ is confirmed to obey
    the standard scaling form $n(s,L)\sim s^{-\tau}\tilde n(s/L^{\df})$, with hyper-scaling relation $\tau=1+d/\df$
	and the spatial dimension $d$. To further characterize the compactness of the FK clusters,
	we measure their graph distances and determine the shortest-path exponents as $\dm(3\rm D)=1.259\,4(2)$
	and $\dm(2\rm D)=1.094\,0(3)$. Further, by excluding all the bridges from the occupied bonds,
	we obtain bridge-free configurations and determine the backbone exponents
	as $\db(3\rm D)=2.167\,3(15)$ and $\db(2\rm D)=1.732\,1(4)$. 
	The estimates of the universal wrapping probabilities for the 3D Ising model and of the geometric critical exponents        $\dm$ and $\db$
	either improve over the existing results or have not been reported yet.
 \end{abstract}
\maketitle

\section{Introduction}

The Ising model~\cite{Ising25} plays an important role
in the study of phase transitions and critical phenomena.
The model exhibits finite-temperature phase transitions in two and more dimensions.
It can be solved exactly for a few two-dimensional lattices~\cite{Onsager44,Baxter-book},
leading to exact values of phase transition points and critical exponents,
which are very often used as benchmarks for new theories and methods.
In three dimensions, since an exact solution of the Ising model is still unavailable,
one usually applies approximation methods or numerical simulations~\cite{FerrenbergXuLandau2018}, 
among which the Monte Carlo (MC) method is probably one of the best.

Most studies of the Ising model focus on the thermodynamic properties.
Quantities of interest include the magnetization, susceptibility, 
energy, heat capacity, and spin-spin correlations etc. 
Under the Fortuin-Kasteleyn (FK) transformation~\cite{FK},
the partition sum of the Ising model can be written as summation over random cluster configurations. 
The FK representation is a key ingredient of the Swendsen-Wang algorithm~\cite{SwendsenWang1987,EdwardsSokal1988}, 
in which non-local updates make the algorithm significantly suppress critical slowing-down.
As geometric objects, the FK clusters also exhibit critical behaviors near the phase transition point.
An example is the wrapping probability~\cite{LanglandsPichetPouliotSaint-Aubin1992,Pinson1994,Arguin2002}, 
which is defined as the probability that there exists a cluster 
which wraps around the periodic boundaries of a finite lattice. 
This quantity takes a universal value at criticality~\cite{LanglandsPichetPouliotSaint-Aubin1992,Pinson1994,ZiffLorenzKleban1999},
and has also been proven to be a good choice in estimating
the critical temperature for continuous phase transitions
due to its impressively small leading finite-size corrections~\cite{WangZhouZhangGaroniDeng2013,NewmanZiff2001},

In this paper, we conduct a Monte Carlo study of the FK clusters of the Ising model on the square (2D) and simple-cubic (3D) lattices.
By performing finite-size scaling (FSS) analysis of the wrapping probabilities, 
we obtain a high-precision estimate of the critical coupling for the 3D Ising model 
as $K_c = 0.221\,654\,631\,(8)$. 
This provides an independent check of the most recent result $K_c=0.221\,654\,626\,(5)$ in Ref.~\cite{FerrenbergXuLandau2018}, 
in which a state-of-the-art method, making use of cross correlations, is applied to reduce statistical errors 
and extensive simulations were carried out up to an impressive linear size $L=1024$.  
The universal values of the wrapping probabilities for the 3D Ising model are determined, which 
have not been reported yet.

Geometric properties of the critical FK clusters are studied.
The probability distribution of the critical largest-cluster size $C_1$ is found  to obey a single-variable 
function as $P(C_1,L){\rm d}C_1=\tilde P(x){\rm d}x$ with $x\equiv C_1/L^{\df}$, 
where the fractal dimension $\df$ is identical to the magnetic renormalization exponent $y_h$. 
The function $\tilde P(x)$ displays a clear bimodal feature in 3D, while in 2D, 
it exhibits an asymmetric peak with a shoulder shape at the smaller-$x$ side. 
By fine-tuning simulations at $L$-dependent coupling $K \to K_c+0^\pm$, 
we find that the two modes correspond to the distinct  asymptotic behaviors approaching from the low- and high-temperature sides.
Then, we consider full FK configurations and measure the cluster number per site $n(s,L)$ of size $s$.
As expected, the standard scaling $n(s,L)\sim s^{-\tau}\tilde n(s/L^{\df})$ is observed both in 2D and 3D, 
and the hyper-scaling relation $\tau=1+d/\df$ is well satisfied ($d$ is the spatial dimension). 
To characterize the compactness of the FK clusters, 
we record their graph distances and determine the shortest-path exponents as $\dm(3\rm D)=1.259\,4(2)$ and $\dm(2\rm D)=1.094\,0(3)$. 
In addition, we classify the occupied bonds into bridges and non-bridges; 
an occupied bond is a bridge iff its deletion leads to the breaking of a FK cluster. 
By excluding all the bridges, we obtain bridge-free configurations and determine the backbone exponents
as $\db(3\rm D)=2.167\,3(15)$ and $\db(2\rm D)=1.732\,1(4)$. 
These estimates of $\dm$ and $\db$ either improve over the existing results or have not been reported yet, to our knowledge. 

The remainder of this paper is organized as follows. 
Section~\ref{simulation} describes the model, simulation and sampled quantities. 
Section~\ref{EstimatingKc} presents the results for the wrapping probabilities of the FK clusters, 
and the estimate of the critical coupling $K_c$ for the 3D Ising model.
Section~\ref{results-at-kc} studies other geometric properties of the FK clusters,
including the probability distribution of the largest-cluster size, the cluster number per site $n(s,L)$,
the graph distances of the FK clusters, the size of the largest cluster in the bridge-free configuration, and the thermodynamic bond densities of various types.
A brief summary and discussion is given in Sec.~\ref{discussions}.
The Appendix collects results on other observables for the Ising model.

\section{Model, simulation and sampled quantities}
\label{simulation}
\subsection{Model}
\label{Model}
We simulate the spin-1/2 Ising model 
on the $L \times L$ square and $L \times L \times L$ simple-cubic 
lattices with periodic boundary conditions, where $L$ is the linear extent.
The spin-1/2 Ising model with a vanishing external field is 
defined by the Hamiltonian
\begin{equation}
\label{Isingaction}
H/k_BT = -K \sum_{\langle \bx \by \rangle} \sigma_{\bx} \sigma_{\by}  \;\;  ,
\end{equation}
where the spins assume values $\sigma_{\bx} \in \{-1, 1 \}$, 
and $\bx =(x_0, x_1)$ or $(x_0,x_1,x_2)$ denotes a vertex with $x_i \in \{1,2,...,L\}$,
$\langle \bx \by \rangle$ sums over all the pairs of nearest neighbors on the lattice. 
Symbols 
$T$, $k_B$, $K$ represent the physical temperature, the Boltzmann constant and 
the reduced coupling, respectively.
In the FK transformation~\cite{FK}, 
for a given lattice $\scrG$ with edge set $\{ \langle \bx \by \rangle \}$, 
the clusters are formed by putting a bond on each edge with probability
$p=(1-e^{-2K})\delta_{\sigma_{\bx} \sigma_{\by}}$.
A cluster is defined as a connected component consisting of vertices and bonds.
Then the partition function can be written as
\begin{equation}
Z_{\rm RC}(v) = 
\sum_{\scrA \subseteq \scrG} v^{\scrN_{\rm bond}} q^{\scrN _{\rm c}} \; ,
	\;\;\; (v= e^{2K}-1)
\label{zrc}
\end{equation}
where $q$ accounts for the statistical weight for each FK cluster, 
$\scrN _{\rm bond}$ and $\scrN _{\rm c}$ are the number of occupied bonds and clusters, respectively.
The summation is over all subgraphs $\scrA$ of the lattice $\scrG$. 
The system is referred to be the random-cluster (RC) model.
The Ising model has $q=2$.

\subsection{Simulation and sampled quantities}
\label{algorithm}
In simulating the Ising model, we employ the Wolff cluster flipping algorithm~\cite{Wolff1989}
and the Swendsen-Wang algorithm~\cite{SwendsenWang1987}.
The latter is used to generate FK clusters over the whole lattice.
The occupied bonds on a complete FK configuration can be classified into bridges and non-bridges
~\cite{Xu14, HuangHouWangZiffDeng2018}. 
A bridge bond is an occupied bond whose deletion would break a cluster.
We delete all bridges to produce a bridge-free configuration.
The corresponding processes are described in Ref.~\cite{Xu14, HuangHouWangZiffDeng2018}, 
which we skip hereby.
Our simulation in 3D is up to $L=512$.
For $L=512$, $384$, $256$ and $192$,
the numbers of samples are about $4 \times 10^6$, $1.5 \times 10^7$, $1.3\times 10^8$ 
and $1.3 \times 10^8$, respectively. 
For each $L \leq 128$, no less than $5 \times 10^8$ samples are generated. 
The 2D simulation is up to $L=1024$.
The numbers of samples are about $3.6\times 10^6$ and $10^7$ for $L=1024$ and $768$, 
respectively,   no less than $2\times 10^7$ samples for $L=512,384,256,192,128$ and $96$, 
and around $10^8$ for each $L\leq 64$.
For a configuration, we sample the following observables:
 \begin{itemize}
 \item The indicators $\scrR^{(x)}$, $\scrR^{(y)}$, and $\scrR^{(z)}$, for the event that
  a cluster wraps around the lattice in the $x$, $y$, or $z$ directions, respectively.

 \item The size of the largest cluster $\scrC_1$.

 \item The size of the largest cluster on the bridge-free configuration $\scrC_{1,\rm{bf}}$. 

 \item An observable $\scrS :=\max\limits_{C}\,\max\limits_{y\in C}\,d(x_C,y)$ 
	is used to determine the shortest-path exponent. Here $d(x,y)$ denotes the graph distance 
	from vertex $x$ to vertex $y$, and $x_C$ is the vertex in cluster $C$
	with the smallest vertex label, according to some fixed (but arbitrary) vertex labeling.

 \item The numbers $\scrN_{\rm b}$, $\scrN_{\rm j}$, $\scrN_{\rm n}$ of branch, junction and non-bridge bonds, respectively. The bridge bond is a junction bond if neither of the two resulting clusters is a tree; otherwise, it is a branch bond~\cite{Xu14, HuangHouWangZiffDeng2018}. 
 \item The square $\scrM^2$ and the fourth power $\scrM^4$ of the magnetization density $\scrM$,
where $\scrM$ is defined as $\scrM=\frac{1}{L^d} \displaystyle{\sum_{x}} \sigma_x$ with $d$ 
the spatial dimension.

 \end{itemize}

 From these observables we calculate the following quantities:
 \begin{itemize}

 \item The wrapping probabilities
 \begin{equation}
   \begin{aligned}
     R^{(x)} = & \langle \scrR^{(x)} \rangle = \langle \scrR^{(y)} \rangle = \langle \scrR^{(z)} \rangle \;, \\
     R^{(2)} = & \langle \scrR^{(x)} \scrR^{(y)}\rangle = \langle \scrR^{(x)} \scrR^{(z)}\rangle =\langle \scrR^{(y)} \scrR^{(z)}\rangle\;, \\
     R^{(3)} = & \langle \scrR^{(x)} \scrR^{(y)}\scrR^{(z)}\rangle\;.
   \end{aligned}
 \end{equation}
 Here $R^{(x)}$, $R^{(2)}$ and $R^{(3)}$ give the probabilities that a winding exists in the $x$ direction,
 in two of the three possible directions, and simultaneously in the three directions, respectively.
 At $K_c$, these wrapping probabilities take non-zero universal values in the thermodynamic limit $L \rightarrow \infty$.

 \item The mean size of the largest cluster $C_1 = \langle \scrC_1 \rangle$,
       which scales as $C_1 \sim L^{\df}$ at $K_c$, with $\df$ the fractal dimension of the FK clusters.

 \item The mean size of the largest cluster 
$C_{1,\rm{bf}} = \langle \scrC_{1,\rm{bf}} \rangle$ in the bridge-free 
configuration, which scales as $C_{1,\rm{bf}} \sim L^{\db}$ at $K_c$, with $\db$ the backbone dimension.

 \item The mean shortest-path distance $S=\langle \scrS \rangle$,
 which scales $S \sim L^{\dm}$ at $K_c$, with $\dm$ the shortest-path fractal dimension.

 \item The number densities $\rho_{\rm b} = \langle \scrN_{\rm b} \rangle/L^d$, 
$\rho_{\rm j} = \langle \scrN_{\rm j} \rangle/V$ and $\rho_{\rm n} = \langle \scrN_{\rm n} \rangle/V$ of the branch, junction and non-bridge bonds, respectively. The leading scaling terms of these bond densities are proportional to $L^{y_t-d}$.

 \item The Binder cumulant
   \begin{equation}
     \begin{aligned}
     Q_m  =& \frac{\langle \scrM^4 \rangle}{\langle \scrM^2 \rangle^2}\;,\;\;\;
     \end{aligned}
     \label{eq:Q}
   \end{equation}

 \end{itemize}

In addition, we record the statistics of the cluster number per site $n(s,L)$ of size $s$, and the 
probability distribution $P(C_1,L){\rm d}C_1$ for the largest-cluster size $C_1$.

For computational efficiency, we use standard re-weighting 
method~\cite{MungerNovotny1991} to obtain the expectations of the wrapping 
probabilities and the Binder cumulant for multiple values of $K$ around $K_c$.

\begin{figure}
\centering
\includegraphics[scale=0.80]{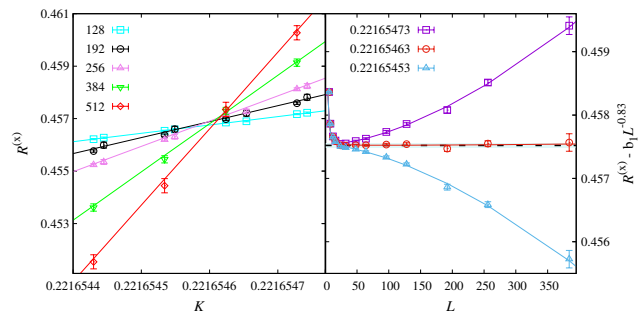}
\caption{Plots of $R^{(x)}$ vs $K$ for different system sizes $L$ (left) and
  $R^{(x)}(K,L)-b_1L^{-0.83}$ vs $L$ for fixed values of $K$(right) for the 3D Ising model. The value of $b_1$ is taken from Table~\ref{Tab:fit-Ising}.
}
\label{Fig:Rx-Ising}
\end{figure}

 \begin{table*}
 \begin{center}
 \caption{Fits of the wrapping probabilities $R^{(x)}, R^{(2)}, R^{(3)}$ and the Binder cumulant 
 $Q_m$ for the 3D Ising model. `Obs.' is the abbreviation of the phrase `observables'.}
 \label{Tab:fit-Ising}
 \scalebox{1.0}{
 \begin{tabular}[t]{|l|l|l|l|l|l|l|l|l|l|l|}
 \hline 
 Obs.                          & $L_{\rm min}$ & $\chi^2/$DF  & $K_c$                 & $y_t$      & $\scrO_c$       & $q_1$         & $b_1$           & $y_i$          & $b_2$        &  $b_3$  \\
 \hline 
 {\multirow{6}{*}{$R^{(x)}$}}  & 12            & 86.1/156     & 0.221\,654\,633(3)    & 1.60(2)    & 0.457\,59(4)    & $-1.4(1)$     &$-0.036(2)$      & $-0.80(2)$     & ~~0.009(9)   & ~~~- \\
                               & 16            & 75.7/140     & 0.221\,654\,631(4)    & 1.59(2)    & 0.457\,53(6)    & $-1.4(1)$     &$-0.040(5)$      & $-0.84(4)$     & ~~0.04(3)    & ~~~- \\
                               & 24            & 71.5/124     & 0.221\,654\,629(4)    & 1.59(2)    & 0.457\,49(9)    & $-1.4(2)$     &$-0.05(2)$       & $-0.88(8)$     & ~~0.08(8)    & ~~~- \\
                               & 12            & 89.4/157     & 0.221\,654\,629(2)    & 1.60(2)    & 0.457\,529(7)   & $-1.38(10)$   &$-0.038\,9(2)$   & $-0.83 $       & ~~0.024(2)   & ~~~- \\
                               & 16            & 75.8/141     & 0.221\,654\,631(2)    & 1.59(2)    & 0.457\,54(1)    & $-1.4(2)$     &$-0.039\,3(3)$   & $-0.83 $       & ~~0.032(4)   & ~~~- \\
                               & 24            & 72.0/125     & 0.221\,654\,631(3)    & 1.59(2)    & 0.457\,55(2)    & $-1.4(2)$     &$-0.039\,4(5)$   & $-0.83 $       & ~~0.03(2)    & ~~~- \\
 \hline
 {\multirow{6}{*}{$R^{(2)}$}}  & 12            & 91.9/134     & 0.221\,654\,633(3)    & 1.60(2)    & 0.332\,01(4)    & $-1.3(1)$     &$-0.093(3)$      & $-0.874(9)$    & $-0.25(1)$   & ~~~- \\
                               & 16            & 73.8/121     & 0.221\,654\,629(4)    & 1.59(2)    & 0.331\,93(6)    & $-1.4(1)$     &$-0.101(6)$      & $-0.90(2)$     & $-0.21(3)$   & ~~~- \\
                               & 24            & 71.2/108     & 0.221\,654\,628(4)    & 1.59(2)    & 0.331\,90(9)    & $-1.4(2)$     &$-0.11(2)$       & $-0.92(4)$     & $-0.17(9)$   & ~~~- \\
                               & 12            & 98.6/134     & 0.221\,654\,637(3)    & 1.60(2)    & 0.332\,12(2)    & $-1.3(1)$     &$-0.080\,3(5)$   & $-0.83 $       & $-0.37(2)$   & 0.45(11) \\
                               & 16            & 75.9/121     & 0.221\,654\,633(3)    & 1.59(2)    & 0.332\,07(3)    & $-1.4(2)$     &$-0.078(1)$      & $-0.83 $       & $-0.47(5)$   & 1.5(4)   \\
                               & 24            & 72.1/108     & 0.221\,654\,631(4)    & 1.59(2)    & 0.332\,04(4)    & $-1.4(2)$     &$-0.076(2)$      & $-0.83 $       & $-0.6(2)$    & 3(2)     \\
 \hline
 {\multirow{6}{*}{$R^{(3)}$}}  & 12            & 110.4/134    & 0.221\,654\,634(3)    & 1.60(2)    & 0.267\,25(4)    & $-1.3(1)$     &$-0.117(3)$      & $-0.885(8)$    & $-0.32(2)$   & ~~~- \\
                               & 16            & 87.9/121     & 0.221\,654\,629(4)    & 1.59(2)    & 0.267\,14(5)    & $-1.3(1)$     &$-0.131(7)$      & $-0.92(2)$     & $-0.25(3)$   & ~~~- \\
                               & 24            & 85.9/108     & 0.221\,654\,629(5)    & 1.59(2)    & 0.267\,13(9)    & $-1.3(2)$     &$-0.13(2)$       & $-0.92(4)$     & $-0.24(9)$   & ~~~- \\
                               & 16            & 86.0/135     & 0.221\,654\,635(3)    & 1.59(2)    & 0.267\,34(3)    & $-1.3(2)$     &$-0.094(1)$      & $-0.83 $       & $-0.66(5)$   & 2.3(4) \\
                               & 24            & 80.1/122     & 0.221\,654\,633(4)    & 1.59(2)    & 0.267\,30(4)    & $-1.3(2)$     &$-0.092(2)$      & $-0.83 $       & $-0.8(2)$    & 4(2)   \\   
                               & 32            & 76.6/109     & 0.221\,654\,628(5)    & 1.58(2)    & 0.267\,19(8)    & $-1.4(2)$     &$-0.085(5)$      & $-0.83 $       & $-1.5(5)$    & 16(8)  \\                            
 \hline
{\multirow{6}{*}{$Q_m$}}       & 12            & 83.3/135     & 0.221\,654\,623(4)    & 1.59(2)    & 1.603\,53(8)    &~~$2.1(2)$     &$-0.271(5)$      & $-0.860(7)$    & $-0.25(3)$   & ~~~- \\
                               & 16            & 80.9/122     & 0.221\,654\,624(5)    & 1.59(2)    & 1.603\,5(2)     &~~$2.1(3)$     &$-0.276(10)$     & $-0.87(2)$     & $-0.22(6)$   & ~~~- \\
                               & 24            & 77.5/109     & 0.221\,654\,623(6)    & 1.59(2)    & 1.603\,6(2)     &~~$2.1(3)$     &$-0.27(3)$       & $-0.85(3)$     & $-0.3(2)$    & ~~~- \\
                               & 12            & 86.0/135     & 0.221\,654\,617(3)    & 1.59(2)    & 1.603\,76(3)    &~~$2.1(3)$     &$-0.245(1)$      & $-0.83 $       & $-0.55(2)$   & 1.0(3) \\
                               & 16            & 80.1/122     & 0.221\,654\,621(4)    & 1.59(2)    & 1.603\,68(5)    &~~$2.1(3)$     &$-0.241(2)$      & $-0.83 $       & $-0.49(4)$   & 2.5(8) \\
                               & 24            & 76.6/109     & 0.221\,654\,622(5)    & 1.59(2)    & 1.603\,65(8)    &~~$2.1(3)$     &$-0.239(4)$      & $-0.83 $       & $-0.65(9)$   & 4(3)   \\
 \hline
 \end{tabular}
}
 \end{center}
 \end{table*}

\section{Wrapping probabilities and critical point in 3D}
\label{EstimatingKc}

The wrapping probability is a universal geometric quantity which reflects topological properties of
the system under study.  This quantity was first introduced for percolation~\cite{LanglandsPichetPouliotSaint-Aubin1992}.  
Later, analytical results of various wrapping probabilities
of percolation clusters on the 2D torus were derived by Pinson~\cite{Pinson1994}, built on works
of Nienhuis~\cite{Nienhuis1984}, di Francesco et al.~\cite{FrancescoSaleurZuber1987}, and Cardy~\cite{Cardy92}.
Arguin extended Pinson's work to the case of 2D RC models with $1 \leq q \leq 4$ and derived
the closed forms of wrapping probabilities in terms of Jacobi $\theta$ functions~\cite{Arguin2002}.
Finite-size corrections of wrapping probabilities for the RC model
in the canonical ensemble (where the total number of the occupied bonds is fixed)  were also studied recently~\cite{HuDeng2015}.
In 3D, there also exist a few studies on the wrapping probabilities~\cite{MartinsPlascak2003, WangZhouZhangGaroniDeng2013}.
Nevertheless, no results concerning the wrapping probability of the FK clusters for the 3D
Ising model have been reported.

In numerical studies of phase transitions, dimensionless quantities like
the Binder cumulant $Q_m$ are known to provide powerful tools for locating
critical points. The wrapping probabilities, topological and dimensionless quantities,  should also provide a useful
method for estimating $K_c$. This is demonstrated in the left plot of Fig.~\ref{Fig:Rx-Ising} 
for the 3D Ising model. The intersections of the $R^{(x)}$ data for different system sizes 
$L$ would give the critical couplings $K_c \approx 0.221\,654\,6$,
with an uncertainty at the seventh decimal place.

In order to estimate $K_c$ more accurately, we resort to the fitting of the data.
Around $K_c$, we perform least-squares fits of the MC data for the wrapping 
probabilities $R^{(x)}$, $R^{(2)}$, $R^{(3)}$ and the Binder cumulant $Q_m$
by the ansatz
\begin{small}
\be
  \begin{aligned}
	\scrO(\epsilon,L) = &\scrO_c + \sum_{k=1}^{2}q_k\epsilon^kL^{ky_t} + c_1\epsilon L^{y_t+y_i} \\
					&+b_1L^{y_i} + b_2L^{-2} + b_3L^{-3}\;\;, \\
  \end{aligned}
\label{eq:pc}
\ee
\end{small}
where $\epsilon = K_c - K$, $\scrO_c$ is a universal constant, $y_t$ is the thermal scaling
exponent and $y_i$ is the leading correction exponent.

As a precaution against correction-to-scaling terms that we have neglected in our chosen ansatz, 
we impose a lower cutoff $L\ge L_{\min}$ on the data points admitted in the fit, 
and systematically study the effect on the $\chi^2$ value when $L_{\min}$ is increased. 
In general, our preferred fit for any given ansatz corresponds to the smallest $L_{\min}$ 
for which $\chi^2$ divided by the number of degrees of freedom (DFs) is $O(1)$, 
and for which subsequent increases in $L_{\min}$ do not cause $\chi^2$ to drop by much more than one unit per degree of freedom.
In the fits with $y_i$ free and $b_3=0$ fixed, 
our results of $y_i$ estimated from $R^{(x)}$ 
is consistent with $y_i \approx -0.83$, as determined 
elsewhere~\cite{Hasenbusch2010a,DengBlote2003}.
In the subsequent fits, we fix $y_i=-0.83$ for all quantities
since in theory $y_i$ should be a universal correction exponent.
In most cases, when performing the fits with $y_i$ fixed, we 
include the correction term $b_3L^{-3}$. However, for $R^{(x)}$ when leave $b_3$ free it can not
be determined, and thus we only use $b_1L^{-0.83}+b_2L^{-2}$ as correction terms.
Table~\ref{Tab:fit-Ising} summarizes the fitting results.

From Table~\ref{Tab:fit-Ising} we observe that 
in comparison with $Q_m$, the wrapping probabilities, 
especially $R^{(x)}$, clearly have smaller amplitudes 
of the leading corrections. 
Due to the weaker corrections, the results of $K_c$ fitted from
the wrapping probabilities have relatively smaller error bars.
We mention that weaker corrections have also been found for wrapping probabilities 
in percolation~\cite{WangZhouZhangGaroniDeng2013}.
From $R^{(x)}$, we estimate $K_c=0.221\,654\,630(6)$ and $R_c^{(x)}=0.457\,5(1)$.
We also obtain $K_c$ and $\scrO_c$ for other observables.
Table~\ref{Tab:fit-Ising} also gives the estimate of the thermal exponent $y_t \approx 1.59$.

After comparing the fits of $K_c$ from various wrapping probabilities,
we present our final estimate as $K_c ({\rm 3D})=0.221\,654\,631(8)$.
The right plot of Fig.~\ref{Fig:Rx-Ising}  demonstrates the values of $K_c$ and $R^{(x)}_c$,
where $R^{(x)}-b_1 L^{-0.83}$ is plotted versus $L$.
The value of the parameter $b_1$ is taken from Table~\ref{Tab:fit-Ising}.
Precisely at $K=K_c$, the $L \to \infty$ data tend to a horizontal line, 
whereas the data with $K \neq K_c$ bend upward or downward.

Our estimate agrees well with the most recent result $0.221\,654\,626(5)$  by Ferrenberg  et~al.~\cite{FerrenbergXuLandau2018} 
within one sigma error bar but with slightly lower precision. 
Since Ref.~\cite{FerrenbergXuLandau2018} used cross correlations to reduce statistical errors 
and carried out simulations up to an impressive linear size $L=1024$, 
our result provides a valuable and independent check.
A previous estimate $0.221\,654\,55(3)$~\cite{DengBlote2003} 
reported by one of our authors and his collaborator is also ruled out.

In Appendix~\ref{sec:estimateyt}, we determine the thermal exponent as $y_t=1.587\,0(5)$
by analyzing the covariance of the wrapping probability
and the energy density, which is also consistent with 
the result $1.587\,5(3)$ in Ref.~\cite{FerrenbergXuLandau2018}.

\section{Geometrical properties of FK clusters at $K_c$}
\label{results-at-kc}
Fixing $K$ at our estimated critical coupling $0.221 \, 654\, 63$ for the simple cubic lattice 
and the exact solution $\ln{(1+\sqrt{2})} / 2 \approx 0.440 \, 686\, 79$ for the square lattice, 
we analyze geometrical quantities defined in Sec.\ref{simulation}.
These include the size of the largest cluster $C_1$, 
the shortest-path distance $S$, 
the size of the largest cluster on the bridge-free configuration $C_{1,{\rm bf}}$. 
These analyses allow us estimate the fractal dimension $\df$, the shortest-path
fractal dimension $\dm$, the backbone fractal dimension $\db$.
In addition, we study the cluster-size distribution
and the probability distribution of the size of the largest cluster. 

\subsection{Fractal dimension $\df$ and the probability distribution of the largest-cluster size}
In order to estimate $\df$, we fit the MC data of $C_1$
to the following equation 
$\df$,
 \be
 \scrA = L^{y_\scrA} ( a_0 + b_1L^{y_1} + b_2L^{y_2})\;.
 \label{eq:A}
 \ee
For the 3D Ising model, when we perform the fit with $y_2=-2$ fixed and $y_1$ free,
we observe that $y_1 \approx -0.83$. To reduce one fitting parameter,
in the subsequent fit we fix $y_1=-0.83$ and $y_2=-2$.
The fitting results are shown in Table~\ref{Tab:fit-C1}.
We also try the fit using $b_1L^{-1}+b_2L^{-2}$ as correction terms
for 2D and 3D, in that case $b_2$ cannot be determined
and the corresponding results are not shown in the table.

\begin{table}
 \caption{Fits of $C_1$ for the 3D and 2D Ising models.}
 \scalebox{0.75}{
 \begin{tabular}{|l|l|l|l|l|l|l|l|}
 \hline
                             & $L_{\rm min}$ & $\chi^2$/DF & $\df$         & $a_0$         & $b_1$           & $y_1$       &  $b_2$ \\
 \hline
    {\multirow{5}{*}{3D}}    &    8          & 6.2/8       & 2.481\,7(3)   & 1.107(2)      &$-0.096(10)$     & $-0.78(7)$  & $-0.30(4)$   \\
                             &   12          & 6.2/7       & 2.481\,7(4)   & 1.107(3)      &$-0.10(3)$       & $-0.79(12)$ & $-0.29(10)$  \\
                             &   12          & 6.3/8       & 2.481\,82(6)  & 1.106\,0(4)   &$-0.106(3)$      & $-0.83$     & $-0.25(3)$   \\
                             &   16          & 5.6/7       & 2.481\,78(8)  & 1.106\,3(5)   &$-0.109(5)$      & $-0.83$     & $-0.21(5)$   \\
                             &   24          & 5.2/6       & 2.481\,84(13) & 1.105\,9(9)   &$-0.104(9)$      & $-0.83$     & $-0.3(2)$    \\
 \hline
   {\multirow{5}{*}{2D}}     &    8          & 14.0/11     & 1.875\,01(2)  & 1.007\,0(1)   &$-0.039(7)$      & $-1.6(1)$   & ~~~-   \\
                             &   12          & 14.0/10     & 1.875\,01(3)  & 1.007\,0(2)   &$-0.04(2)$       & $-1.6(3)$   & ~~~-   \\
							 &   16          & 12.3/9      & 1.875\,02(3)  & 1.007\,0(2)   &$-0.12(15)$      & $-2.0(5)$   & ~~~-   \\
                             &   16          & 12.3/10     & 1.875\,024(13)  & 1.006\,91(6)  &$-0.12(1)$     & $-2$        & ~~~-   \\
                             &   24          & 12.2/9      & 1.875\,020(17)  & 1.006\,93(8)  &$-0.13(3)$     & $-2$        & ~~~-   \\
\hline
 \end{tabular}
}
\label{Tab:fit-C1}
\end{table}

\begin{figure}
\centering
\includegraphics[scale=0.65]{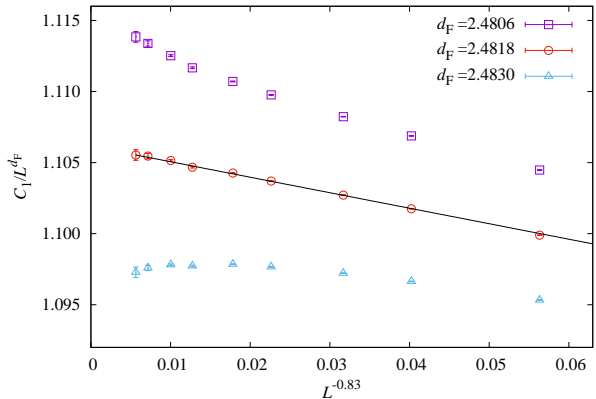}
\caption{Plots of $C_1/L^{d_F}$ vs $L^{-0.83}$ for the critical 3D Ising model.} 
\label{Fig:df}
\end{figure}

Comparing these fits, we determine the fractal dimensions $\df=2.481\,8(4)$ for 3D 
and $1.875\,01(4)$ for 2D.
The latter agrees with the exact value $y_h=15/8$~\cite{Yang52,Nienhuis87}.
In Fig.~\ref{Fig:df}, we plot $C_1/ L^ {\df}$ versus $L^{-0.83}$
using three different values of $\df$ for the 3D Ising model:
namely our estimate, as well as the estimate plus or minus three standard deviations.
As $L$ increases, the data with $\df=2.480\,6$ and $\df=2.483\,0$
bend upward and downward, respectively, while the data with $\df=2.481\,8$
are consistent with a straight line. 
This illustrates the reliability of our estimate.

\begin{figure}
\centering
\includegraphics[scale=0.65]{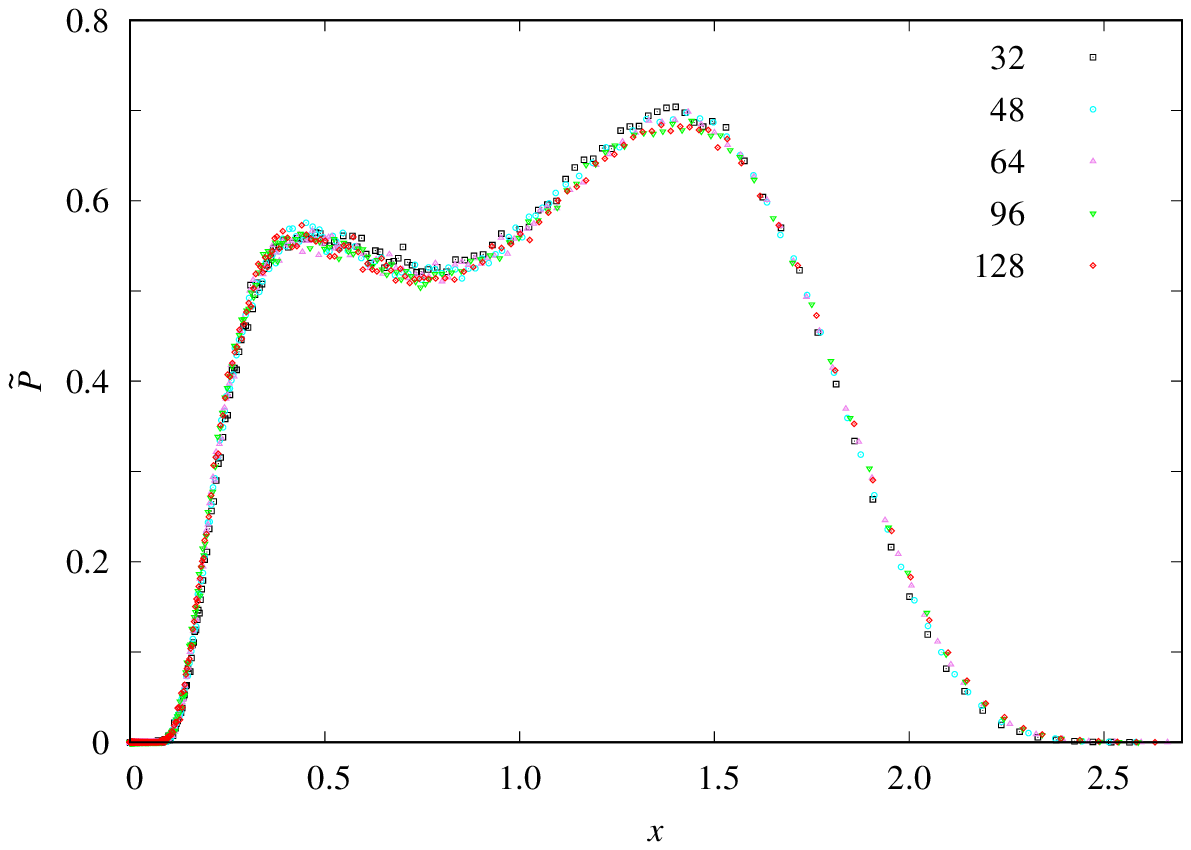} \\
\includegraphics[scale=0.65]{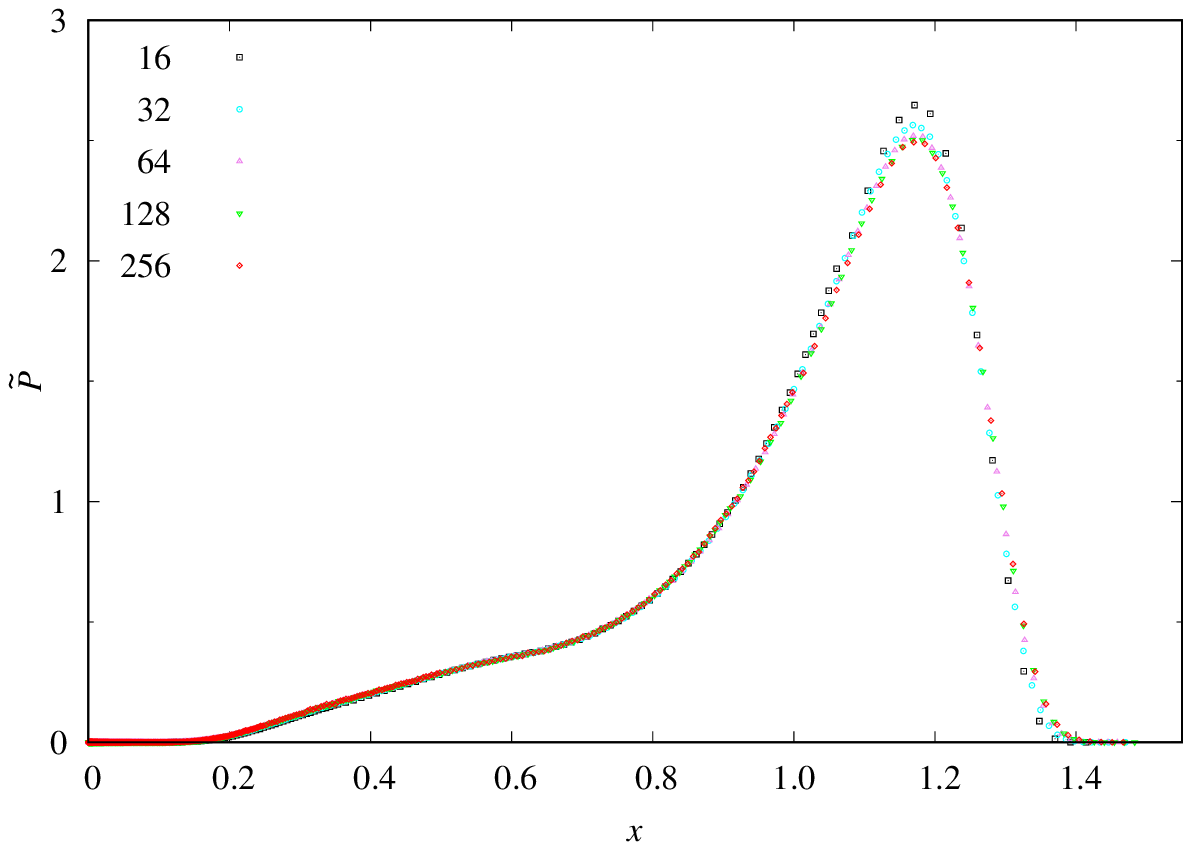} \\
\caption{Probability density distribution of the largest-cluster size 
	for 3D (top) and 2D (bottom) at $K_c$, with $x \equiv C_1/L^{\df}$.} 
\label{Fig:sdC1}
\end{figure}

We also study the probability distribution 
$P(C_1,L) {\rm d} C_1$ of the size of the largest cluster $ C_1$.
In the MC simulation, $P(C_1,L){\rm d} C_1$ is measured by the fraction of 
the number of the configurations on which the size of the largest cluster
lies between $C_1 \sim C_1+{\rm d} C_1$.
According to finite-size scaling theory,
we expect that $P(C_1,L){\rm d}C_1$ 
can be expressed as a single-variable function $\tilde P(x){\rm d}x$, 
with $x\equiv C_1/L^{\df}$.
This is well confirmed by Fig.~\ref{Fig:sdC1}, where the data for different system sizes 
collapse on top of each other.
Interestingly, we see that the scaling function
$\tilde P(x)$ exhibits a bimodal structure in 3D, 
and a single peak with a wide shoulder shape in the small-$x$ side in 2D.

\begin{figure}
\centering
\includegraphics[scale=0.65]{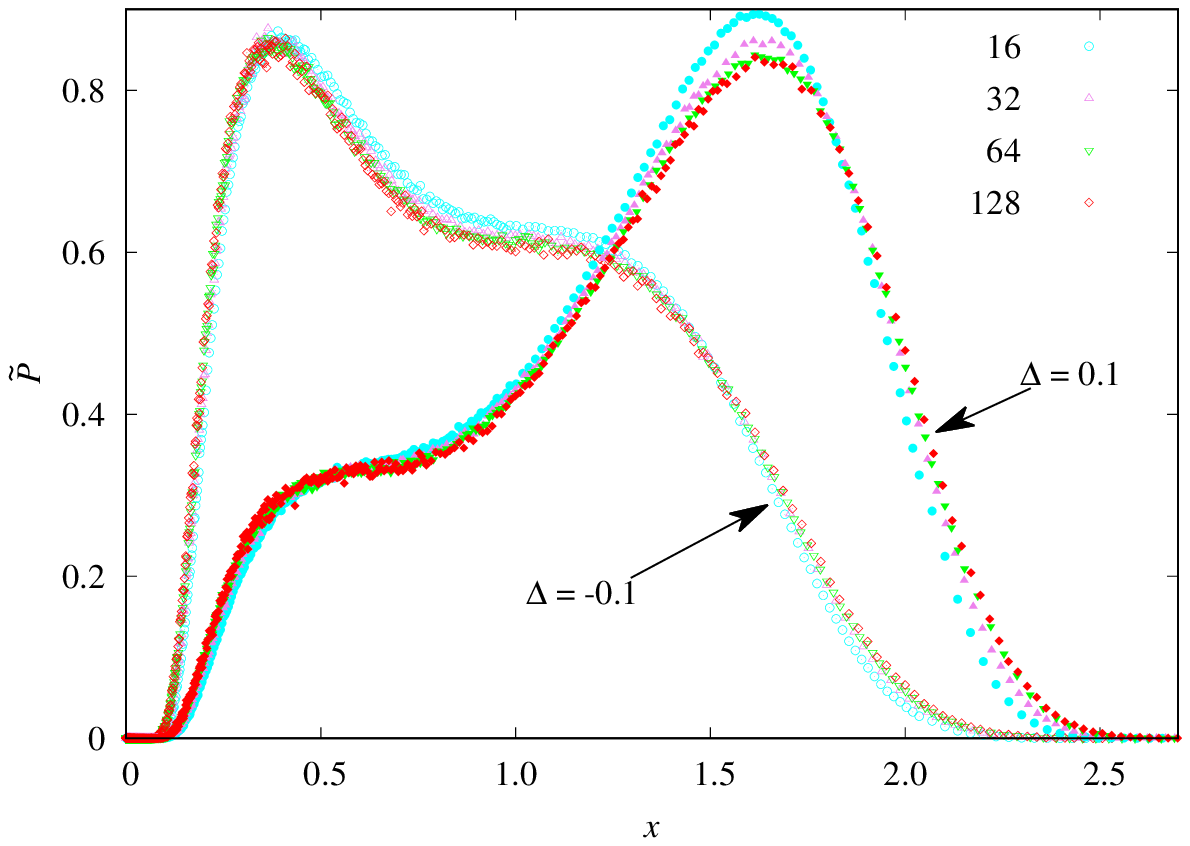} \\
\includegraphics[scale=0.65]{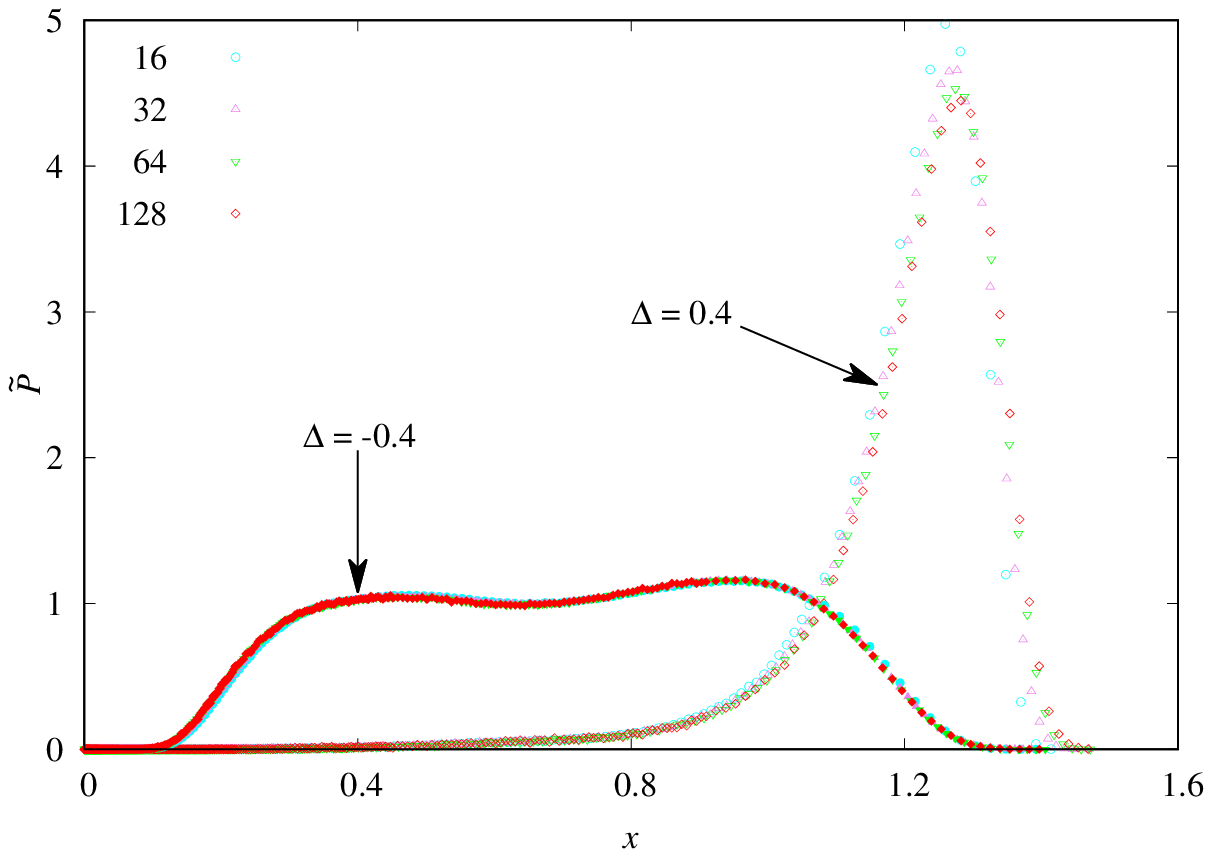} \\
\caption{Probability density distribution of the largest-cluster size
	for 3D (top) with $K=K_c\pm \Delta L^{-y_t}$ and 2D (bottom) with $K=K_c\pm \Delta L^{-y_t}$.}
\label{Fig:sdC1pm}
\end{figure}

To understand the bimodal structure in 3D, we explore $\tilde P(x)$ in the critical window $\Delta \equiv L^{y_t} (K-K_c)$, 
with $\Delta$ a finite constant. An example with $\Delta = \pm 0.1$ is shown in the top plot of Fig.~\ref{Fig:sdC1pm},
where the distributions become to have a single peak with a wide shoulder shape.
Therefore, it is reasonable to assume that the asymptotical peak locations $x_{\rm max} (\Delta \to 0^{\pm} )$ are actually different.
From the similarity between  the 3D  low-temperature distribution with $\Delta=0.1$ (the top plot of Fig.~\ref{Fig:sdC1pm}) 
and the 2D critical one (the bottom plot of Fig.~\ref{Fig:sdC1}),
we expect that a bimodal distribution would appear for 2D in the high-temperature region with $\Delta < 0$. 
The 2D results with $\Delta = \pm 0.4$ are shown in the bottom plot of Fig.~\ref{Fig:sdC1pm}, confirming our expectation.

\subsection{Cluster-size distribution}
\begin{figure}
\centering
\includegraphics[scale=0.65]{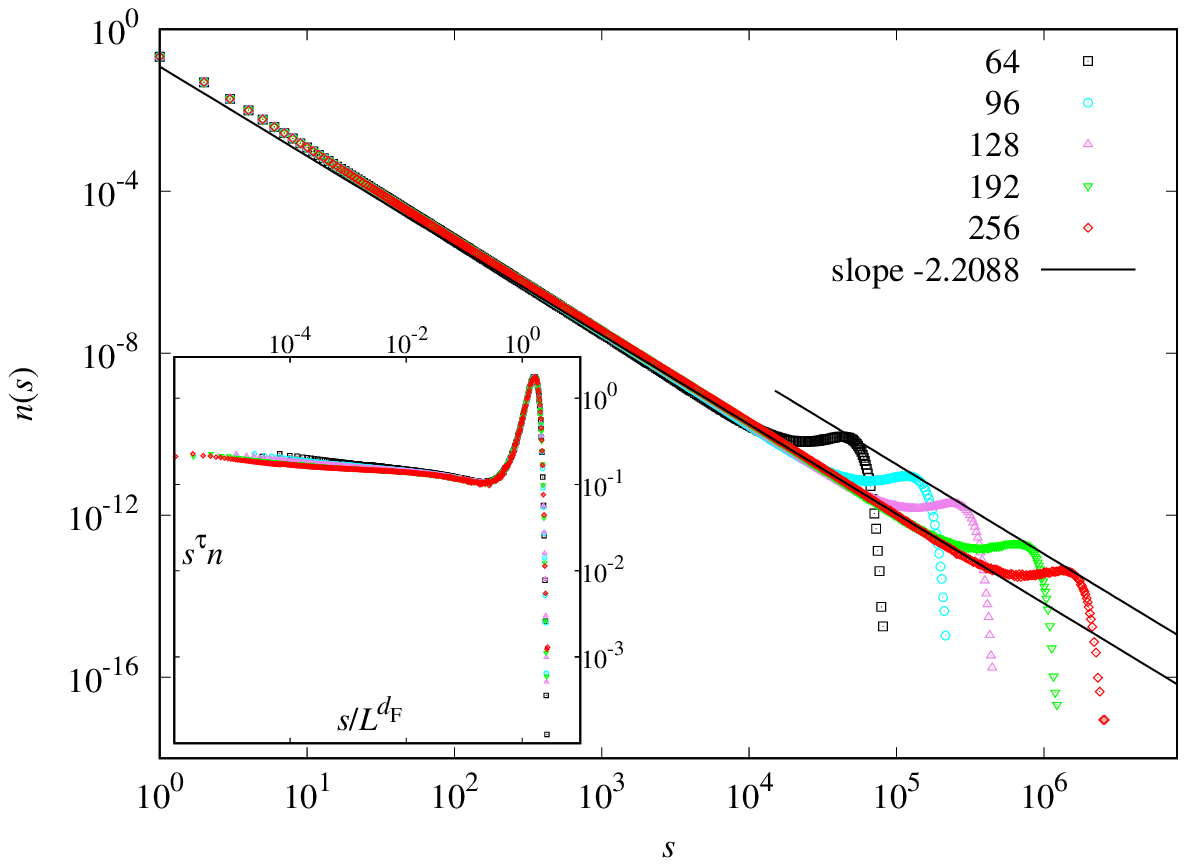} \\
\includegraphics[scale=0.65]{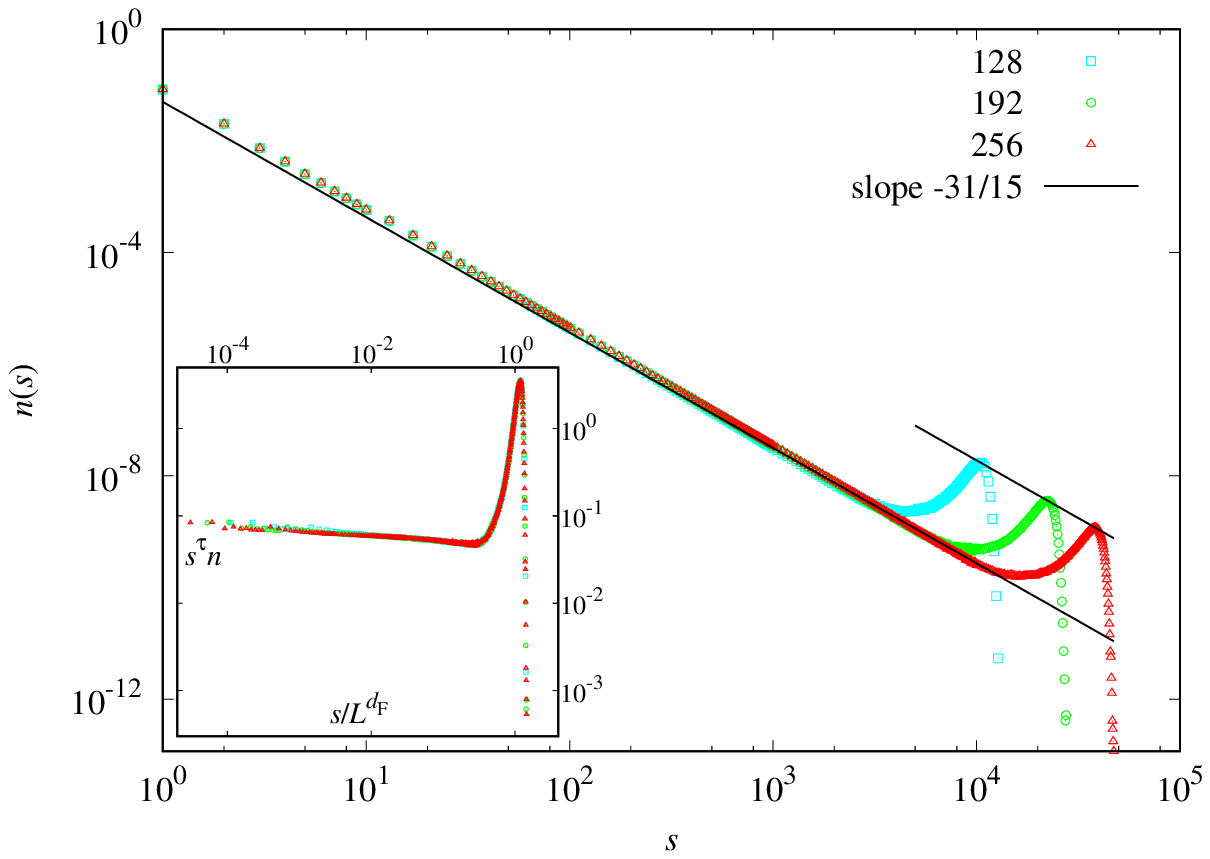} \\
\caption{Cluster-size distribution for FK clusters for  3D (top)
and  2D (bottom) at criticality. In both cases, the insets
show $s^{\tau}n(s)$ vs $s/L^{\df}$.}
\label{Fig:sd}
\end{figure}

We consider the critical cluster-number density $n(s,L)$ of size $s$,
of which the scaling behavior is expected to follow
 \be
  n(s,L) = s^{-\tau}{\tilde n}(s/L^{\df}) \; ,
 \label{eq:distribution}
 \ee
where $\tau=1+d/{\df}$ is Fisher exponent and
${\tilde n}(x)$ with $x\equiv s/L^{\df}$ is a universal scaling function.
From ${\df} \approx 2.4818$ (3D) or $15/8$ (2D),
one has $\tau \approx 2.2088$ (3D) or $31/15$ (2D), respectively.
In the main plots of Fig.~\ref{Fig:sd}, we show a log-log plot of $n(s)$ versus
$s$ for $L=64,96,128,192$ and $256$ for the 3D Ising model,
and for $L=128,192$ and $256$ for the 2D Ising model.
The straight lines with slope $-2.2088$ (3D) and $31/15$ (2D) are drawn for comparison
with the MC data. In these plots, we observe clearly the power-law behaviors 
$n(s,\infty) \sim s^{-2.2088}$ (3D) and $n(s,\infty) \sim s^{-31/15}$ (2D), respectively.

In order to display the universal scaling function ${\tilde n}(x)$,
we further plot $s^{2.2088}n(s,L)$ versus $s/L^{2.4818}$ (3D)
and $s^{31/15}n(s,L)$ versus $s/L^{15/8}$ (2D) for several system sizes,
and show them in the insets of Fig.~\ref{Fig:sd}.
We find a good collapse of those curves for different system sizes,
which provides strong numerical evidence for the conjectured scaling
Eq.(\ref{eq:distribution}).

\subsection{Shortest-path fractal dimension $\dm$}
We estimate the shortest-path fractal dimension $\dm$ for the 3D and 2D
Ising models by studying the shortest-path distance $S$.
The MC data for $S$ are fitted to Eq.~\eqref{eq:A} with the exponent
$y_\scrA$ being replaced by $\dm$. 
For the 3D Ising model,
with $y_2=-3$ and $y_1$ being free, we obtain $y_1=-1.8(2)$, much smaller than $y_i \approx -0.83$ from 
the leading irrelevant thermal scaling field.
On this basis, we further perform the fit with $y_1=-1.8$ and $y_2=-3$ fixed.
For the 2D Ising model, setting $b_2=0$ and $y_1$ free,
$y_1$ cannot be determined by our MC data. We then try the fit with $y_1=-2$
fixed and find $b_1$ consistent with zero. 
On this basis, we perform the fit with $b_1=0$ and $b_2=0$.
Thus, $S$ suffers rather small finite-size corrections both in 2D and 3D.

From these fits, we estimate $\dm = 1.259\,4(2)$ (3D) and $1.094\,0(3)$ (2D),
respectively.
As far as we know, the shortest-path fractal dimension of the 3D
Ising FK clusters has not been estimated.
The 2D result improves over the previous reported value $\dm=1.095\,5(10)$~\cite{DengZhang2010}.

\begin{figure}
\centering
\includegraphics[scale=0.65]{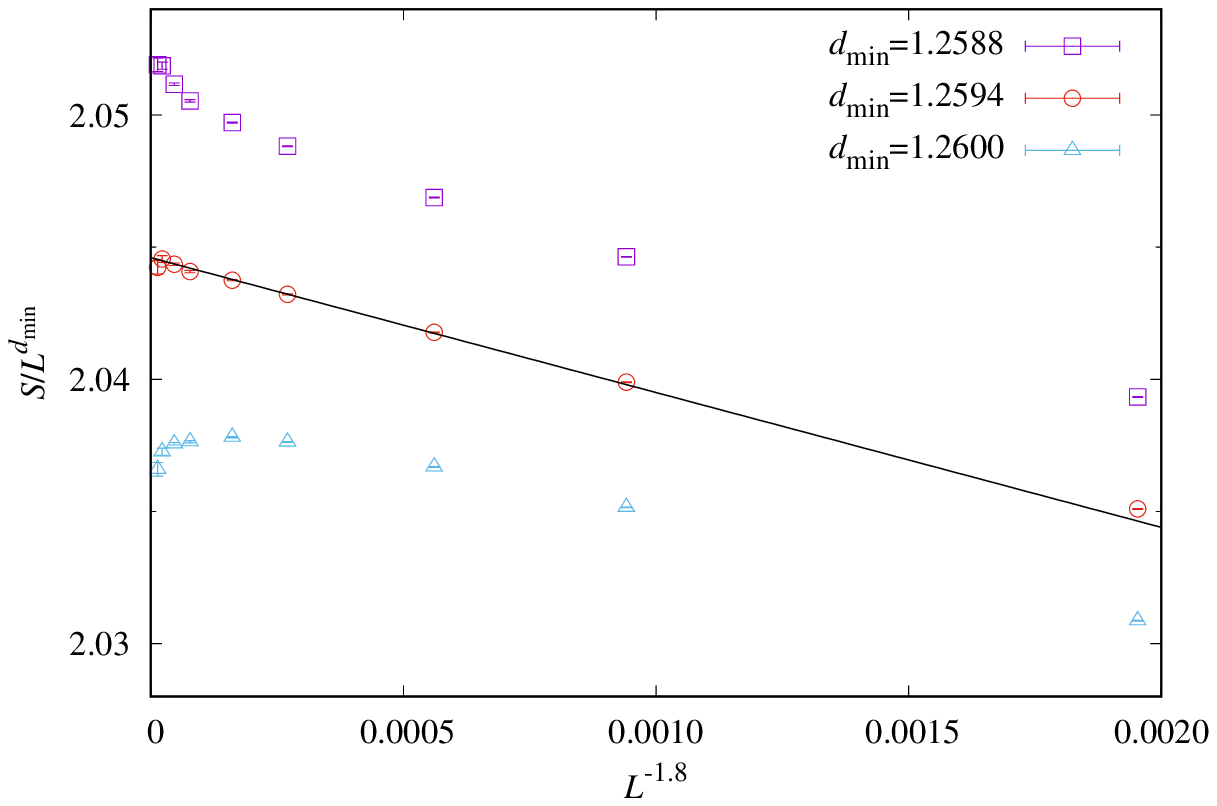} \\
\includegraphics[scale=0.65]{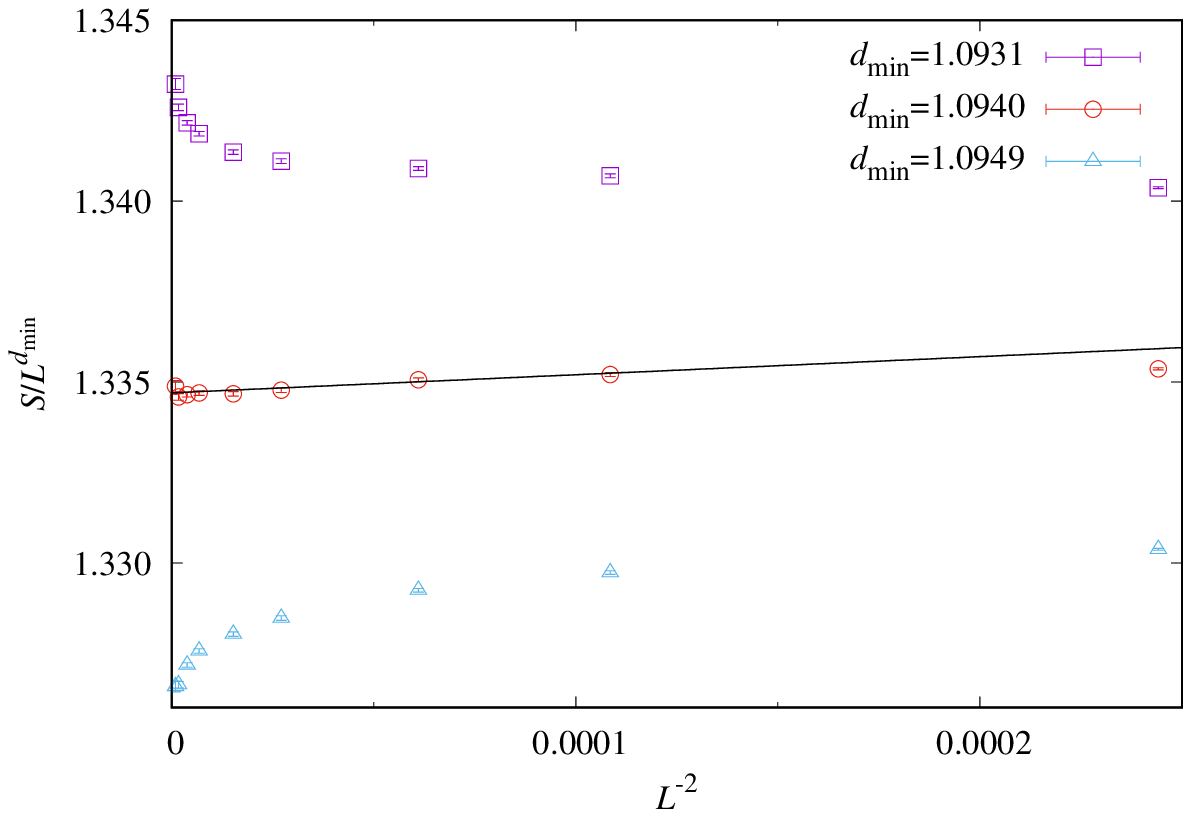}
	\caption{Plots of $S/L^{d_{\rm {min}}}$ vs $L^{-1.8}$ or 
	$L^{-2}$ for the critical 3D (top) or 2D Ising model (bottom), respectively.}
	\label{Fig:dmin}
\end{figure}

To illustrate our estimate, Fig.~\ref{Fig:dmin} shows a
plot of $S/L^{\dm}$ versus $L^{-1.8}$ (3D) and a plot of
$S/L^{\dm}$ versus $L^{-2}$ (2D) at three different $\dm$ values.
In both cases, using the estimated values of $\dm$ produces a straight line,
in contrast the other two curves bend upward or downward for large $L$.
The figure suggests that the true value of $\dm$ does indeed
lie within 3$\sigma$ of our estimate.

\begin{table}
	\caption{Fits of $S$ for the 3D and 2D Ising models.}
 \scalebox{0.85}{
 \begin{tabular}[t]{|l|l|l|l|l|l|l|l|}
 \hline
                         &  $L_{\rm min}$  & $\chi^2$/DF & $\dm$               &  $a_0$         &  $b_1$        & $y_1 $         & $b_2$      \\
 \hline
  {\multirow{5}{*}{3D}}  &  16             &   9.3/6     & 1.259\,34(5)        & 2.045\,3(5)    &  $-5.0(8)$    & $-1.78(5)$     & $27(5)$    \\
                         &  24             &   6.2/5     & 1.259\,42(6)        & 2.044\,3(7)    &  $-9(4)$      & $-1.9(1)$      & $60(24)$   \\
                         &  16             &   9.5/7     & 1.259\,36(2)        & 2.045\,0(2)    &  $-5.38(4)$   & $-1.8$         & $29.5(7)$  \\
                         &  24             &   8.0/6     & 1.259\,34(3)        & 2.045\,2(3)    &  $-5.48(9)$   & $-1.8$         & $33(3)$    \\
                         &  32             &   7.0/5     & 1.259\,37(4)        & 2.044\,9(4)    &  $-5.3(2)$    & $-1.8$         & $25(9)$    \\
 \hline
  {\multirow{4}{*}{2D}}
						 &  96             &    6.7/5    & 1.093\,99(6)       & 1.334\,7(5)    &~~$ 5(2)$  	    & $-2$           & ~~~-       \\
						 & 128             &    2.9/4    & 1.094\,09(8)       & 1.333\,9(6)    &~~$10(3)$  		& $-2$           & ~~~-       \\
						 & 192             &    2.9/3    & 1.094\,08(12)      & 1.334(1)       &~~$9(9)$  		& $-2$           & ~~~-       \\
                         & 192             &    4.2/4    & 1.093\,96(5)       & 1.335\,0(4)    &  ~~~- 		    & ~~~-           & ~~~-       \\
                         & 256             &    3.0/3    & 1.094\,00(6)       & 1.334\,7(5)    &  ~~~-   	    & ~~~-           & ~~~-       \\
 \hline
 \end{tabular}
 }
 \label{Tab:fit-S}
 \end{table}

\subsection{Backbone fractal dimension $\db$}
In order to estimate the backbone fractal dimension $d_{\rm B}$ for the
3D and 2D Ising models, we fit the MC data of $C_{1,\rm{bf}}$ to Eq.(\ref{eq:A})
with $y_\scrA$ being replaced by $d_{\rm B}$.
For the 3D Ising model, in the fit with $y_2=-2$ fixed and $y_1$ free,
we observe that $y_1 \approx -0.83$. On this basis we fix $y_1=-0.83$ and $y_2=-2$.
For the 2D Ising model, when set $b_2=0$ and leave $y_1$ free,
we find that $y_1 \approx -0.67$, suggesting rather strong finite-size corrections. 

\begin{figure}
\centering
\includegraphics[scale=0.65]{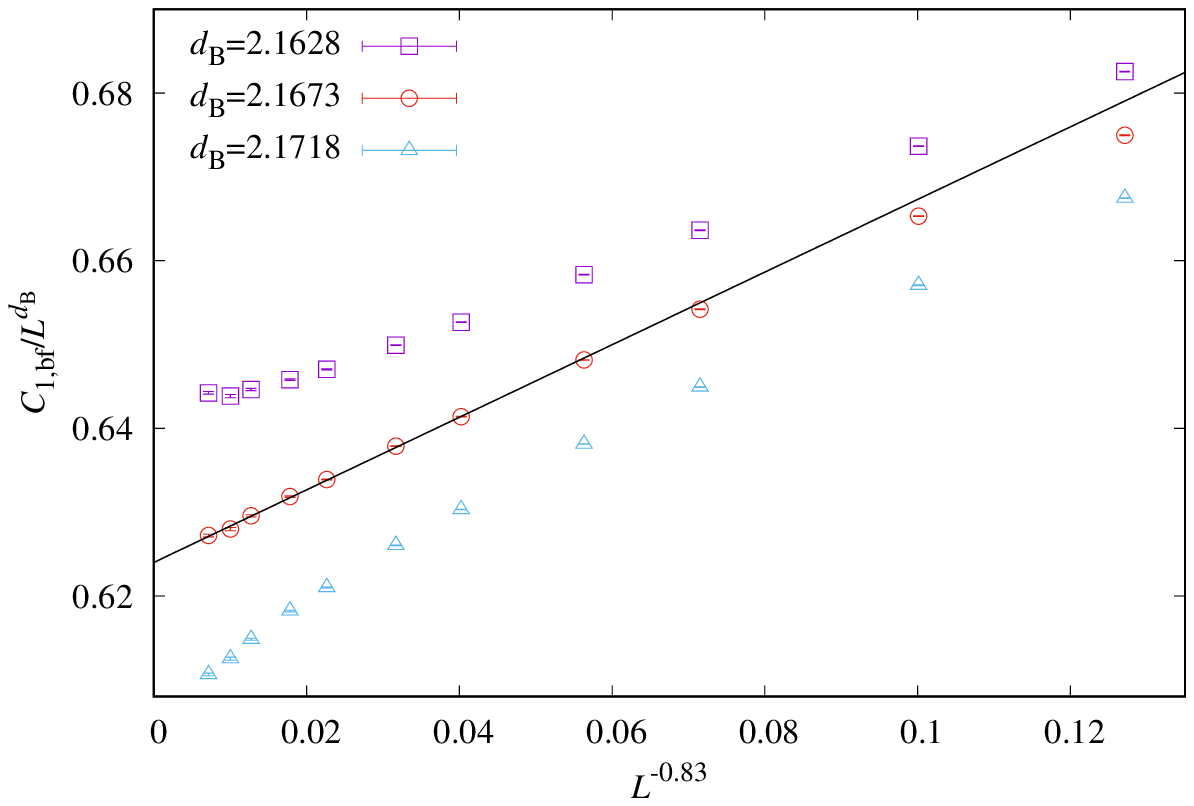} \\
\includegraphics[scale=0.65]{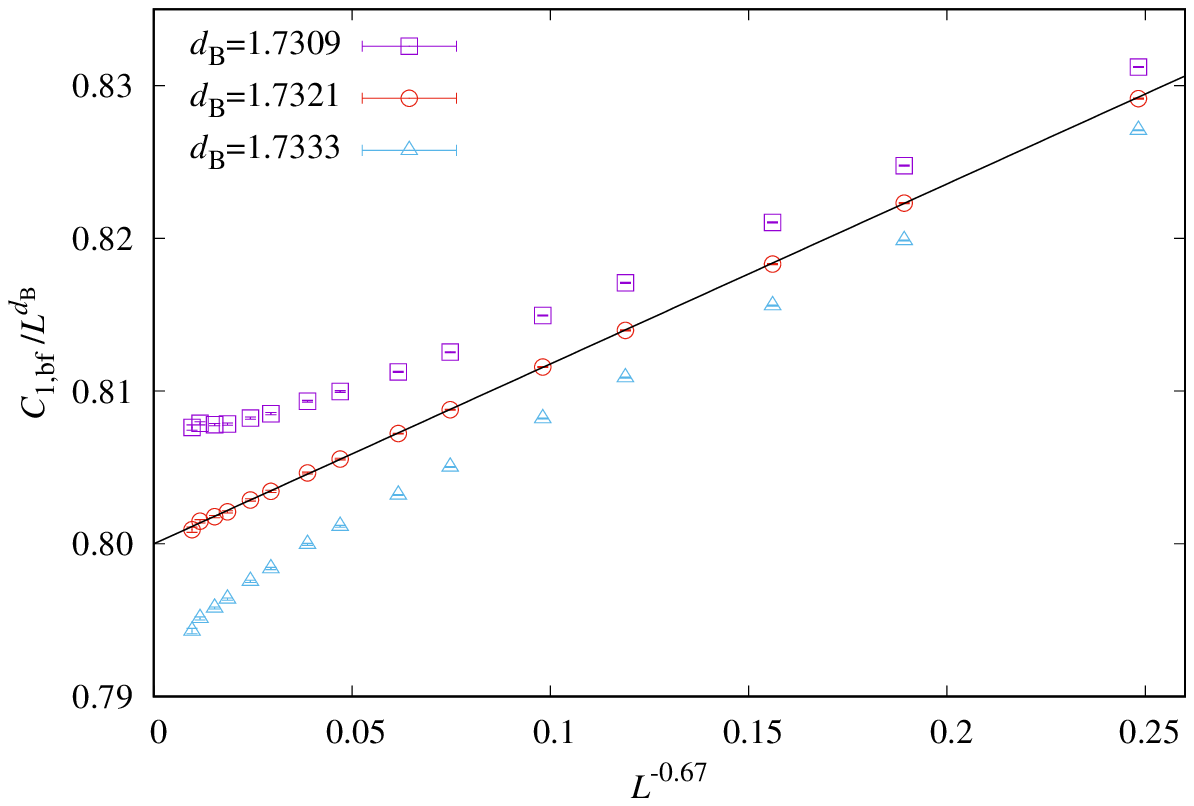}
\caption{Plots of $C_{1,\rm{bf}}/L^{d_{\rm B}}$ vs $L^{-0.83}$
or vs $L^{-0.67}$ for the critical 3D (top) or 2D (bottom)
Ising model, respectively.}
\label{Fig:C1bf}
\end{figure}

The fitting results are shown in Table~\ref{Tab:fit-C1BF}.
Comparing these fits, we estimate the backbone fractal dimension as $\db=2.167\,3(15)$ (3D)
and $1.732\,1(4)$ (2D), respectively. 
The 3D result improves over the previous reported value
$\db(3\rm D)=2.171(4)$~\cite{DengBlote2004-1}, 
and the 2D result rules out the previous estimate
$\db(2\rm D)=1.730\,4(3)$~\cite{DengBloteNienhuis2004-3}.
These previous reported values were estimated by studying the scaling behavior of
the probability that a pair of lattice sites at a distance $r$
are connected by at least two mutually independent paths.
Similarly, by using three different $d_{\rm B}$ values,
in Fig.~{\ref{Fig:C1bf}}
we plot $C_{1,\rm{bf}}/L^{d_{\rm B}}$ versus $L^{-0.83}$ (3D)
or $L^{-0.67}$ (2D), illustrating the reliability of our estimate of $\db$.

 \begin{table}
	 \caption{Fits of $C_{1,\rm{bf}}$ for the 3D and 2D Ising models.}
 \scalebox{0.81}{
 \begin{tabular}[t]{|l|l|l|l|l|l|l|l|}
 \hline
                            &      $L_{\rm min}$  & $\chi^2$/DF & $d_{\rm{B}}$      & $a_0$             & $b_1$        & $y_1$         & $b_2$  \\
 \hline
  {\multirow{5}{*}{3D}}     &       8             &   8.9/6     & 2.168\,0(7)       & $0.621(3)$        & 0.40(2)      & $-0.78(3)$    & $-0.50(8)$  \\
                            &      12             &   6.9/5     & 2.166\,9(9)       & $0.626(4)$        & 0.46(6)      & $-0.85(6)$    & $-0.8(3)$   \\
                            &      12             &   7.0/6     & 2.167\,3(2)       & $0.624\,2(8)$     & 0.439(6)     & $-0.83$       & $-0.71(5)$  \\
                            &      16             &   6.0/5     & 2.167\,1(3)       & $0.615(1)$        & 0.433(9)     & $-0.83$       & $-0.6(1)$ \\
                            &      24             &   4.8/4     & 2.167\,4(5)       & $0.624(2)$        & 0.45(2)      & $-0.83$       & $-0.9(3)$   \\
 \hline
  {\multirow{6}{*}{2D}}     &       8             &  12.7/11    & 1.732\,14(13)     & $0.7997(7)$       & 0.116\,8(9)  & $-0.66(2)$  & ~~~-        \\
                            &      12             &  10.5/10    & 1.732\,0(2)       & $0.8008(10)$      & 0.119(2)     & $-0.69(2)$    & ~~~-        \\
                            &      16             &   8.6/9     & 1.732\,2(3)       & $0.7996(14)$      & 0.116(3)     & $-0.66(3)$    & ~~~-        \\
                            &       8             &  13.1/12    & 1.732\,07(4)      & $0.800\,1(2)$     & 0.117\,2(4)  & $-0.67$        & ~~~-        \\
                            &      12             &  11.2/11    & 1.732\,10(4)      & $0.800\,0(2)$     & 0.117\,8(6)  & $-0.67$        & ~~~-        \\
                            &      16             &   8.8/10    & 1.732\,06(5)      & $0.800\,2(3)$     & 0.116\,9(9)  & $-0.67$        & ~~~-        \\
 \hline
 \end{tabular}
 }
 \label{Tab:fit-C1BF}
 \end{table}

 \subsection{Bond densities $\rho_b$, $\rho_j$ and $\rho_n$}
 In order to estimate the critical bond densities for branch,
 junction and non-bridge bonds for the 3D and 2D Ising models,
 we fit the MC data of $\rho_b$, $\rho_j$ and $\rho_n$ to the ansatz
 　 \be
 　 \rho = \rho_{0} +L^{y_t-d}(a+bL^{y_1}) \;
 　 \label{eq:rho}
 　 \ee
 with $y_t-d=-1.413$ (3D) and $-1$ (2D) fixed.
 For the 3D Ising model, in the fits with $y_1$ free,　we observe that $y_1 \approx -1.2$ for $\rho_b$,
 however, for $\rho_j$ and $\rho_n$, we can not obtain stable fitting results.
 On this basis, in the subsequent fits we fix $y_1=-1.2$.
 For the 2D Ising model, the correction exponent $y_1$ for
 $\rho_b$, $\rho_j$ and $\rho_n$ can not be determined by
 our MC data when leaving it free.
 We then try the fits with fixed $y_1=-1$ or $-2$ respectively.
 in both cases, $b_1$ is found to be consistent with zero. On this basis,
 we perform the fit with fixed $b_1=0$.
 The fitting results are shown in Table~\ref{Tab:fit-rhobjn-Ising}.
 \begin{table}
 \caption{Fits of the bond densities $\rho_b$, $\rho_j$ and $\rho_n$
 for the 3D and 2D Ising models.}
  \scalebox{0.82}{
 \begin{tabular}[t]{|l|l|l|l|l|l|l|l|}
 \hline
                        &            & $L_{\rm min}$  & $\chi^2$/DF & $\rho_{0}$          &  $a$             & $b$              & $y_1$     \\
 \hline
 {\multirow{12}{*}{3D}}  &            &  8             & 6.2/8       & 0.176\,526\,50(5)   & $-0.136\,9(1)$   &~~$0.025(6)$      & $-1.2(2)$ \\
                       &            & 12             & 4.3/7       & 0.176\,526\,47(5)   & $-0.136\,8(1)$   &~~$0.05(4)$       & $-1.5(3)$ \\
                        & $\rho_b$   & 16             & 3.9/6       & 0.176\,526\,49(6)   & $-0.136\,9(2)$   &~~$0.03(3)$       & $-1.2(5)$ \\
                        &            & 12             & 5.5/8       & 0.176\,526\,50(4)   & $-0.136\,92(5)$  &~~$0.028(2)$      & $-1.2$    \\
                       &            & 16             & 3.9/7       & 0.176\,526\,49(4)   & $-0.136\,88(6)$  &~~$0.026(3)$      & $-1.2$    \\
                        &            & 24             & 3.6/6       & 0.176\,526\,48(4)   & $-0.136\,85(8)$  &~~$0.023(5)$      & $-1.2$    \\
                        \cline{2-8}
                        &            & 24             & 6.9/6       & 0.010\,298\,17(2)   & $-0.070\,29(3)$  &~~$0.008(2)$      & $-1.2$    \\
                        & $\rho_j$   & 32             & 5.8/5       & 0.010\,298\,18(2)   & $-0.070\,32(4)$  &~~$0.011(4)$      & $-1.2$    \\
                        &            & 48             & 4.8/4       & 0.010\,298\,19(3)   & $-0.070\,37(8)$  &~~$0.018(9)$      & $-1.2$    \\
                        \cline{2-8}
                        &            & 12             & 5.3/8       & 0.051\,342\,34(9)   &~~$0.340\,1(1)$   & $-0.052(4)$      & $-1.2$    \\
                        & $\rho_n$   & 16             & 4.9/7       & 0.051\,342\,35(9)   &~~$0.340\,1(2)$   & $-0.049(6)$      & $-1.2$    \\
                        &            & 24             & 4.9/6       & 0.051\,342\,35(10)  &~~$0.340\,1(2)$   & $-0.05(1)$       & $-1.2$    \\
 \hline
 {\multirow{12}{*}{2D}}  &            & 24             & 7.4/9       & 0.183\,250\,4(3)    & $-0.109\,68(9)$  &~~$0.004(3)$      & $-1$      \\
                        & $\rho_b$   & 24             & 6.4/9       & 0.183\,250\,3(3)    & $-0.109\,64(6)$  &~~$0.09(6)$       & $-2$      \\
                        &            & 24             & 9.6/10      & 0.183\,250\,0(3)    & $-0.109\,56(4)$  &~~~-              &~~~-       \\
                        &            & 32             & 4.7/9       & 0.183\,250\,2(3)    & $-0.109\,60(4)$  &~~~-              &~~~-       \\
                        \cline{2-8}
						&            & 16             & 8.7/10      & 0.023\,856\,2(2)    & $-0.110\,52(4)$  & $-0.0001(8)$     & $-1$      \\
                        & $\rho_j$   & 16             & 8.7/10      & 0.023\,856\,2(2)    & $-0.110\,52(3)$  &~~$0.001(9)$      & $-2$      \\
                        &            & 16             & 8.7/11      & 0.023\,857\,2(2)    & $-0.110\,52(2)$  &~~~-              &~~~-       \\
                        &            & 24             & 8.6/10      & 0.023\,856\,2(2)    & $-0.110\,52(2)$  &~~~-              &~~~-       \\
                        \cline{2-8}
                        &            & 24             & 7.5/9       & 0.292\,893\,6(8)    &~~$0.311\,4(2)$   & $-0.005(6)$      & $-1$      \\
                        & $\rho_n$   & 24             & 7.0/9       & 0.292\,893\,7(7)    &~~$0.311\,4(1)$   & $-0.1(1)$        & $-2$      \\
                        &            & 24             & 8.2/10      & 0.292\,894\,0(6)    &~~$0.311\,25(7)$  &~~~-              &~~~-       \\
                       &            & 32             & 5.2/9       & 0.292\,893\,7(6)    &~~$0.311\,33(9)$  &~~~-              &~~~-       \\
 \hline
 \end{tabular}
 }
 \label{Tab:fit-rhobjn-Ising}
 \end{table}

After comparing various fits, we obtain the critical thermodynamic bond densities of various types, 
including the branch bonds $\rho_{b,0} ({\rm 3D})=0.176\,526\,5(1)$ and $\rho_{b,0}({\rm 2D}) =0.183\,250\,2(5)$,
the junction bonds $\rho_{j,0} ({\rm 3D}) =0.010\,298\,2(1)$ and $\rho_{j,0} ({\rm 2D})=0.023\,856\,2(2)$, 
as well as the non-bridge bonds $\rho_{n,0} ({\rm 3D}) =0.051\,342\,35(10)$ and $\rho_{n,0} ({\rm 2D})=0.292\,893\,7(9)$.
The non-bridge density in 2D is consistent well with the exact result $0.292\,893\,219$~\cite{Eren16,Hu14}.
Among all the occupied bonds, the fraction of the branch, junction
and non-bridge bonds are $74.12 \%$, $4.32 \%$ and $21.56 \%$ for the 3D Ising
model, $36.65 \%$, $4.77 \%$ and $58.58 \%$ for the 2D Ising model, respectively.
This suggests that as the spatial dimension $d$ increases, the critical FK clusters become more and more dentritic.

 \begin{table}[t]
 \caption{Summary of the estimates of the critical exponents and 
	 the universal critical wrapping probabilities.}
 \label{Tab:summary}
 \scalebox{0.95}{
 \begin{tabular}{|c|c c|c c|}
 \hline
	 &\multicolumn{2}{c|}{2D}				&\multicolumn{2}{c|}{3D}  \\
	 $$ 		  &Present		  &Previous							&Present	     &Previous \\
 \hline
	 $R_c^{(x)}$  &~~- 		  &$0.627\,138\,794$~\cite{LanglandsPichetPouliotSaint-Aubin1992}	&$0.457\,5(1)$   &~~-	\\
	 $R_c^{(2)}$  &~~- 		  &$0.480\,701\,867$~\cite{LanglandsPichetPouliotSaint-Aubin1992}	&$0.332\,0(2)$   &~~-	\\
	 $R_c^{(3)}$  &~~- 		  &~~-							&$0.267\,2(2)$   &~~-	\\
	 \hline
	 $\df$		  &$1.875\,01(5)$ &$15/8$~\cite{Yang52, Nienhuis87}	&$2.481\,8(4)$   & $2.481\,6(1)$~\cite{DengBlote2003} \\
	 $\dm$		  &$1.094\,0(3)$  &$1.095\,5(10)$~\cite{DengZhang2010}			&$1.259\,4(2)$	 &~~-	\\
	 $\db$		  &$1.732\,1(4)$  &$1.730\,4(3)$~\cite{DengBloteNienhuis2004-3}	&$2.169\,2(15)$  &$2.171(4)$~\cite{DengBlote2004-1}\\
 \hline
 \end{tabular}
 }
 \end{table}

\section{Summary and Discussion}
\label{discussions}
In this work, we investigate the Ising model from the perspective of the geometric properties of the FK clusters. 
We find that the wrapping probabilities, a kind of topological quantities, 
suffer less from finite-size corrections near the critical point, and thus 
they provide a powerful tool for locating the critical point. 
This leads to a high-precision estimate for the 3D Ising model $K_c=0.221\,654\,631(8)$, 
a competing result with the most recent one $0.221\,654\,626(5)$~\cite{FerrenbergXuLandau2018}.
The probability distribution is observed to follow a single-variable function 
$P(C_1,L){\rm d}C_1 \equiv \tilde P(x)dx$, with $x \equiv C_1/L^{\df}$. 
The scaling function $\tilde P(x)$ displays a very rich behavior within the scaling 
window $\Delta \equiv L^{y_t} (K-K_c)$, with a finite constant $\Delta$, 
including a bimodal feature. 
We also study other quantities that characterize the geometric ``compactness" 
of the critical FK clusters. In particular, we determine the shortest-path fractal dimension $\dm(3\rm D)=1.259\,4(2)$
and $\dm(2\rm D) = 1.094\,0(3)$ from the graph distances and 
the backbone fractal dimension $d_{\rm B}(3\rm D)=2.169\,2(15)$ and $d_{\rm B}(2\rm D) = 1.732\,1(4)$. 
A brief summary is given in Table~\ref{Tab:summary}.

These results, together with the thermodynamic bond densities of various types, 
suggest that as the spatial dimension increases, the critical FK clusters become more and more dentritic. 
The FK representation of the Ising model provides much richer critical behaviors than the spin representation, 
which are not well understood yet. 
For instance, even in 2D, it remains to be an open question whether the shortest-path and backbone dimensions 
take some fractional numbers, and if so, what their values are. 

Finally, we note that recent developments of the logarithmic conformal field theory have provided some new insights 
for non-local geometric correlation functions for the critical Potts model,
 including the 3D percolation model~\cite{Tan2018,CouvreurJacobsen2017}.
Another recent development is the conformal-bootstrap program,
which leads to a very high-precision estimate of critical exponents for local operators for the 3D Ising model~\cite{El-Showk2012,Poland2018}.
Our work might provide a solid numerical test ground for some fantastic theoretical developments in future.

\section{Acknowledgments}
This work was supported by the National Science Fund for Distinguished
Young Scholars (NSFDYS) under Grant No.~11625522 (Y.J.D), the National Natural
Science Foundation of China (NSFC) under Grant No.~11405039 (J.F.W), 
the Fundamental Research Fund for the Central Universities under
Grant No.~J2014HGBZ0124 (J.F.W),
and by Anhui University under Start-up Grant No.~J01006187 (H.H).

\appendix

\section{Other quantities}
\label{sec:other-observables}
In addition to those in the main text, 
we have also considered several other quantities in the Monte Carlo simulations, including
\begin{itemize}
 \item The energy density $\scrE$ and its square $\scrE^2$, where $\scrE$ is defined as 
$\scrE=\frac{1}{d L^d} \displaystyle{\sum_{<xy>}} \sigma_x \sigma_y$. 
 \item The number of clusters $\scrN_{\rm c}$.

 \item The second cluster-size  moments $\scrS_2  = \sum_{k} \scrC_k^2$,
  where the sum runs over all the clusters and $\scrC_k$ denotes the size of the $k$th cluster.
 \item An observable $\scrF :=\frac{1}{dL^d}\displaystyle{\sum_{k=0}^{d-1} |\sum_{\bx}}\sigma_\bx \exp(i \frac{2\pi x_k}{L})|^2$,
 which is the Fourier transform of the correlation function at the lowest nonzero momentum.
\end{itemize}

We measure the following quantities:
\begin{itemize}
 \item The covariance of $\scrR^{(x)}$ and $\scrE$
 \begin{equation}
   g^{(x)}_{ER} = \langle \scrR^{(x)} \scrE \rangle - \langle \scrR^{(x)}\rangle \langle \scrE \rangle \; ,
 \label{eq:g}
 \end{equation}
 which scales as $g^{(x)}_{ER}\sim L^{y_t} $ at $K_c$, with $y_t=1/\nu$ the thermal exponent.
\item The cluster number density $\rho = \langle \scrN_{\rm c} \rangle/L^d$, whose leading scaling term is propotional to $L^{y_t-d}$.

 \item Specific heat $C_e=L^d (\langle \scrE^2\rangle-{\langle \scrE \rangle}^2)$, 
  which scales as $C_e\sim L^{2y_t-d} = L^{\alpha/\nu}$ at $K_c$.
 \item Susceptibility $\chi = \langle \scrS_2 \rangle / L^d$, which scales as
 $\chi \sim L^{2d_F-d} = L^{\gamma/\nu}$ at $K_c$.
 \item The second moment correlation length 
 \begin{equation}
   \xi_{2nd} = \sqrt{\frac{\chi/F-1}{4\sin^2\pi/L}} \; ,
 \label{eq:corlen}
 \end{equation}
 where $F=\langle \scrF \rangle$. 
 At $K_c$, the ratio $\xi_{2nd}/L$ takes an non-zero universal value 
 in the thermodynamic limit $L \rightarrow \infty$.
\end{itemize}

\subsection{Estimating $y_t$}
\label{sec:estimateyt}
We estimate $y_t$ by studying the covariance $g^{(x)}_{ER}$ for the
3D and 2D Ising models at the critical couplings $K=0.221\,654\,63$ (3D) and
$K_c=0.440\,686\,79$ (2D), respectively. The MC data is fitted to
Eq.~(\ref{eq:A}) with $y_\scrA$ being replaced by $y_t$,
We note that, in percolation case~\cite{WangZhouZhangGaroniDeng2013},
similar procedure for estimating $y_t$ has been found preferable to methods,
such as that employed in~\cite{DengBlote2005},
in which $y_t$ is estimated by studying how quantities behave in the
neighborhood of the percolation threshold.

For the 3D Ising model, in the fit with $b_2=0$ fixed and $y_1$ free,
we find $y_1 \approx -0.83$. We then perform the fit with $y_1=-0.83$ and
$y_2=-2$ fixed.
For the 2D Ising model, when leave $b_2=0$ fixed and $y_1$ free,
we determine $y_1 \approx -0.46$. 
The fitting results are shown in Table~\ref{Tab:fit-g}.
 \begin{table}
 \caption{Fits of $g_{ER}^{(x)}$ for the 3D and 2D Ising models.}
 \scalebox{0.82}{
 \begin{tabular}[t]{|l|l|l|l|l|l|l|l|}
 \hline
                       & $L_{\rm min}$          & $\chi^2$/DF       &  $y_t$             & $a_0$          & $b_1$            & $y_1$            & $b_2$  \\
 \hline
 \multirow{6}{*}{3D}   & 12                     & 6.5/8             & 1.586\,7(3)        & 0.492\,2(9)    & $-0.651(3)$      & $-0.815(5)$      &  ~~~~-  \\
                       & 16                     & 6.0/7             & 1.586\,5(4)        & 0.493(2)       & $-0.648(6)$      & $-0.811(7)$      &  ~~~~-  \\
                       & 24                     & 4.8/6             & 1.587\,0(6)        & 0.491(2)       & $-0.66(2)$       & $-0.82(2)$       &  ~~~~-  \\
                       & 12                     & 7.4/8             & 1.587\,2(2)        & 0.490\,4(4)    & $-0.672(3)$      & $-0.83$          &  0.09(3)  \\
                       & 16                     & 5.0/7             & 1.587\,0(2)        & 0.490\,9(5)    & $-0.677(5)$      & $-0.83$          &  0.15(5)  \\
                       & 24                     & 4.7/6             & 1.587\,2(3)        & 0.490\,6(9)    & $-0.673(10)$     & $-0.83$          &  0.1(2) \\
 \hline
 \multirow{6}{*}{2D}   &  6                     &  7.5/12           & 0.999\,8(8)        & 0.428(3)       & $-0.474(2)$      & $-0.456(4)$      &  ~~~~-  \\
                       &  8                     &  7.4/11           & 0.999\,8(10)       & 0.428(4)       & $-0.474(2)$      & $-0.456(6)$      &  ~~~~-  \\
					   & 12                     &  7.0/10           & 1.000\,5(15)       & 0.426(5)       & $-0.473(2)$      & $-0.46(1)$     &  ~~~~-  \\
                       &  8                     &  7.9/12           & 1.000\,4(2)      & 0.425\,9(3)    & $-0.472\,6(8)$   & $-0.46$          &  ~~~~-  \\
                       & 12                     &  7.0/11           & 1.000\,3(2)        & 0.426\,2(5)    & $-0.473(1)$      & $-0.46$          &  ~~~~-  \\
                       & 16                     &  6.9/10           & 1.000\,2(3)        & 0.426\,3(6)    & $-0.474(2)$      & $-0.46$          &  ~~~~-  \\
 \hline
 \end{tabular}
  }
 \label{Tab:fit-g}
 \end{table}

\begin{figure}
\centering
\includegraphics[scale=0.65]{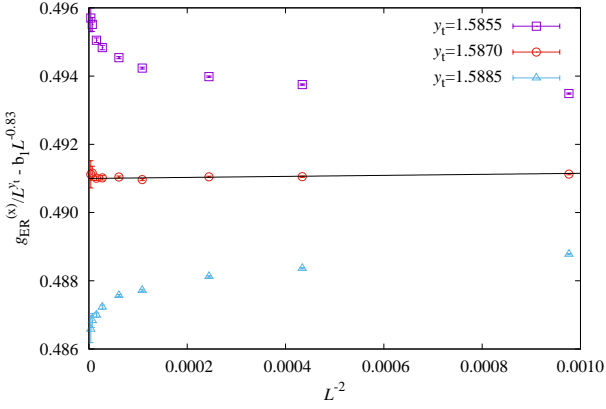}
\caption{Plot of $g^{(x)}_{ER}/L^{y_t} - b_1 L^{-0.83}$
versus $L^{-2}$ for the 3D Ising model, 
using three different values of $y_t$.
The value of $b_1$ is taken from Table~\ref{Tab:fit-g}.}
\label{Fig:yt}
\end{figure}

After comparing various fits, we estimate the thermal scaling exponent
for the 3D and 2D Ising models as $y_t=1.587\,0(5)$ (3D) and $-1.000(1)$ (2D),
respectively.
In order to illustrate our estimate of $y_t$ for the 3D Ising model,
we plot $g^{(x)}_{ER}/L^{y_t} - b_1 L^{-0.83}$
versus $L^{-2}$ using three different values of $y_t$:
our estimate, as well as our estimate plus or minus three standard deviations,
and show them in Fig.~\ref{Fig:yt}.
Using the estimated value of $y_t$ should produce a straight line for large $L$.
In the figure, the data using $y_t=1.585\,5$ and $y_t=1.588\,5$ respectively
bend upward and downward, suggesting that the true value of $y_t$ does indeed lie within
$3\sigma$ of our estimate. The data with $y_t = 1.587\,0$ appear to be consistent with an
asymptotically straight line. For the 2D Isnig model, our estimate of $y_t$ is
consistent well with the analytical result $y_t=1$, as expected.

\subsection{Cluster number density $\rho$}
\label{estimaterho}

At $K=K_c$ and for $L \to \infty$, the FK cluster number density
shall approach to an non-universal (model-dependent) constant $\rho_0$.
We fit the MC data of $\rho$ for the 3D and 2D Ising models 　to~Eq.(\ref{eq:rho}).
 For the 3D Ising model, in the fit with $y_t$ and $y_1$ free, we observe that 
 $y_t=1.587(3)$ which is consistent with our estimated $y_t=1.5870(5)$.
On this basis, we further perform the fit with $y_t=1.587$ fixed and $y_1$ free, and observe that the correction
exponent $y_1\approx -1.47$. To reduce one fitting parameter, we also try the fit with $y_t=1.587$ and $y_1=-1.47$ fixed.
For the 2D Ising model, if letting $y_t=1$ fixed and $y_1$ free, we observe that the correction 
exponent $y_1\approx -1$. To reduce one fitting parameter, we then try the fit with $y_1=-1$ fixed.
The fitting results are shown in Table.~\ref{Tab:fit-rho}.
After comparing various fits, we estimate the critical cluster number densities
as $\rho_0=0.315\,588\,2(2)$ (3D) and $0.128\,679\,6(6)$ (2D), respectively.

\begin{figure}
\centering
\includegraphics[scale=0.8]{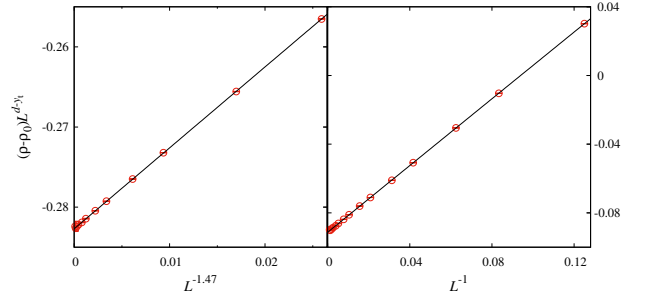}
	\caption{Plots of the cluster number density $(\rho-\rho_0)L^{1.413}$ versus 
	$L^{-1.47}$ for the 3D (left) Ising model 
	and $(\rho-\rho_0)L$ versus $L^{-1}$ for the 2D (right) Ising model at the critical temperature.}
\label{Fig:rhok}
\end{figure}

In Fig.~\ref{Fig:rhok}, we plot $(\rho-\rho_0)L^{d-y_t}$
versus $L^{-1.47}$ (3D) and versus $L^{-1}$ (2D).
In both cases, for large system sizes the data points are arranged in an
straight line, as expected. 
 \begin{table}
 \caption{Fits of the cluster number density $\rho$ for the 3D and 2D Ising models.}
 \scalebox{0.74}{
 \begin{tabular}[t]{|l|l|l|l|l|l|l|l|}
 \hline
						&  $L_{\rm min}$  & $\chi^2$/DF & $\rho_{0}$         & $a$              & $y_t$	 	      &  $b$         & $y_1$         \\
 \hline
{\multirow{7}{*}{3D}}   &  12             & 2.4/7       & 0.315\,588\,26(9)  & $-0.282\,0(9)$   & $1.587\,5(7)$  & $1.04(4)$    & $-1.49(2)$   \\
						&  16             & 1.4/6       & 0.315\,588\,2(1) & $-0.283(2)$      & $1.587(1)$ & $0.96(9)$    & $-1.45(5)$    \\
						&  24             & 1.2/5       & 0.315\,588\,2(2) & $-0.282(3)$      & $1.587(2)$ & $1.1(3)$     & $-1.49(12)$    \\
						&  12             & 3.0/8       & 0.315\,588\,19(4)  & $-0.282\,67(6)$  & $1.587$		  & $1.01(2)$    & $-1.470(7)$       \\
						&  16             & 1.4/7       & 0.315\,588\,21(5)  & $-0.282\,73(7)$  & $1.587$	 	  & $0.98(3)$    & $-1.46(2)$       \\
						&  12             & 3.0/9       & 0.315\,588\,19(3)  & $-0.282\,67(3)$  & $1.587$		  & $1.009(2)$    & $-1.47$       \\
						&  16             & 2.5/8       & 0.315\,588\,18(4)  & $-0.282\,66(3)$  & $1.587$	 	  & $1.007(3)$    & $-1.47$       \\
\hline
{\multirow{6}{*}{2D}}   &   8             & 10.6/11     & 0.128\,679\,5(3)   & $-0.091\,2(1)$   & $1$	 	 	  & $0.974(3)$     & $-1.002(2)$   \\
						&  12             & 10.6/10     & 0.128\,679\,4(4)   & $-0.091\,2(2)$   & $1$	 	      & $0.974(8)$     & $-1.002(4)$   \\
                        &  16             &  7.8/9      & 0.128\,679\,1(4)   & $-0.091\,0(2)$ & $1$  	      & $0.993(14)$    & $-1.009(6)$   \\
                        &   8             & 11.2/12     & 0.128\,679\,6(2)   & $-0.091\,29(4)$  & $1$	 	      & $0.971\,5(5)$  & $-1$          \\
                        &  12             & 10.8/11     & 0.128\,679\,6(3)   & $-0.091\,27(5)$  & $1$	 	      & $0.971\,0(9)$  & $-1$          \\
                        &  16             & 10.3/10     & 0.128\,679\,6(3)   & $-0.091\,30(6)$  & $1$	 	      & $0.972(2)$     & $-1$          \\
 \hline
 \end{tabular}
 }
 \label{Tab:fit-rho}
 \end{table}

\subsection{Specific heat $C_e$}
According to the scaling theory, specific heat at criticality
scales as $C_e \sim L^{\alpha/\nu}$.
In order to fit the MC data of $C_e$,
the fitting ansatz Eq.(\ref{eq:A}) is
reformulated by adding a constant term $c_0$ due to the existence of analytic
background, leading to
 \be
 \scrA = c_0+L^{y_\scrA} ( a_0 + b_1L^{y_1} + b_2L^{y_2})\;,
 \label{eq:Ce}
 \ee
where the exponent $y_\scrA$ stands for $\alpha/\nu$.
For the 3D Ising model,
in the fit with $b_2=0$ fixed and $y_1$ free, we observe that
$y_1 \approx -1.5$. To reduce one fitting parameter,
we perform the subsequent fit with both $y_1=-1.5$ and $b_2=0$ fixed.
Besides, we also perform the fit with both $y_1=-0.83$ and $y_2=2y_1=-1.66$ fixed.

For the 2D Ising model, since $\alpha=0$, the leading scaling
term $L^{y_\scrA}$ changes to $\ln L$. We fit the MC data of $C_e$ to the following equation~\cite{FerdinandFisher1969,Salas2001}
\be
 \scrA =  a_0\ln L + c_0 + b_1 L^{-1} + b_2L^{-2}\;.
 \label{eq:Ce2D}
\ee
In the fitting results $b_2$ is consistent with zero. On this basis, we perform the fit with $b_2=0$ fixed.
The fitting results are reported in Table~\ref{Tab:fit-Ce}.
For 2D Ising model, $a_0=0.6366(5)$ is consistent
with the theoretical value $a_0=2/\pi$~\cite{Salas2001}.

\begin{table}
\caption{Fits of $C_e$ for the 3D and 2D Ising models.}
 \scalebox{0.75}{
 \begin{tabular}[t]{|l|l|l|l|l|l|l|l|l|}
 \hline
					  &   $L_{\rm min}$     & $\chi^2$/DF     & $\alpha/\nu$        & $c_0$         & $a_0$         & $b_1$          & $y_1$      & $b_2$ \\  
\hline
 \multirow{8}{*}{3D}   &    8                & 7.1/8           &$0.169(1)$         & $-3.61(7)$    &~~$4.83(6)$    & $-1.76(10)$    & $-1.49(6)$  &~~~~-  \\
					  &   12                & 7.0/7           &$0.170(2)$         & $-3.6(2)$     &~~$4.8(1)$     & $-1.7(3)$      & $-1.4(2)$    &~~~~-\\
                      &   16                & 4.5/6           &$0.174(5)$         & $-3.3(4)$     &~~$4.6(3)$     & $-1.0(3)$      & $-1.1(3)$ 	  &~~~~-\\
                      &   12                & 7.2/8           &$0.169\,3(6)$      & $-3.62(4)$    &~~$4.84(3)$    & $-1.77(4)$     & $-1.5$    	  &~~~~- \\
                      &   16                & 7.0/7           &$0.169\,1(7)$      & $-3.63(5)$    &~~$4.85(4)$    & $-1.74(8)$     & $-1.5$    	  &~~~~- \\
                      &   24                & 4.7/6           &$0.170(2)$         & $-3.54(8)$    &~~$4.78(6)$    & $-2.0(2)$      & $-1.5$    	  &~~~~- \\
					  &    8				& 6.7/8		      &$0.171(2)$		  & $-3.5(1)$	  &~~$4.73(8)$	  & $-0.24(8)$	   & $-0.83$	  & $-1.7(2)$ \\
					  &   12				& 6.7/7			  &$0.171(3)$		  & $-3.4(2)$	  &~~$4.7(2)$	  & $-0.3(2)$	   & $-0.83$	  & $-1.7(4)$ \\
\hline
\multirow{5}{*}{2D}   &    6                &10.7/12          &~~~~-		        &$0.178(2)$   & $0.6365(4)$  & $-0.22(2)$ 	& $-1$   	  & $0.05(6)$ \\     
					  &    8                &10.1/11          &~~~~-       		    &$0.180(3)$   & $0.6363(5)$  & $-0.24(3)$     & $-1$     	  & $0.1(2)$ \\
					  &    6                &11.3/13          &~~~~-      		    &$0.1772(9)$  & $0.6367(2)$  & $-0.211(4)$    & $-1$        &~~~~-\\  
					  &	   8				&11.2/12		  &~~~~-				&$0.1774(1)$  & $0.6367(3)$  & $-0.212(6)$    & $-1$  	  &~~~~-\\  
					  &	  12				&10.3/11		  &~~~~-				&$0.1785(2)$  & $0.6365(4)$  & $-0.22(1)$	    & $-1$		  &~~~~-    \\
 \hline
 \end{tabular}
 } 
 \label{Tab:fit-Ce}
 \end{table}

\subsection{Susceptibility $\chi$}
We fit the MC data of $\chi$ in 3D and 2D to Eq.(\ref{eq:A}) with 
the exponent $y_\scrA$ replaced by $2\df-d$. 
For the 3D Ising model, in the fit with $y_2=-2$ fixed and $y_1$ free, we observe that $y_1 \approx -0.83$.
To reduce one fitting parameter, in the subsequent fit we fix $y_1=-0.83$ and $y_2=-2$.
For the 2D Ising model, when leave $b_2=0$ fixed and $y_1$ free, we determine $y_1\approx -2$. On this 
basis, we perform the fit with $y_1=-2$ and $b_2=0$ fixed.
The fitting results are shown in Table~\ref{Tab:fit-S2}.

\begin{figure}
\centering
\includegraphics[scale=0.65]{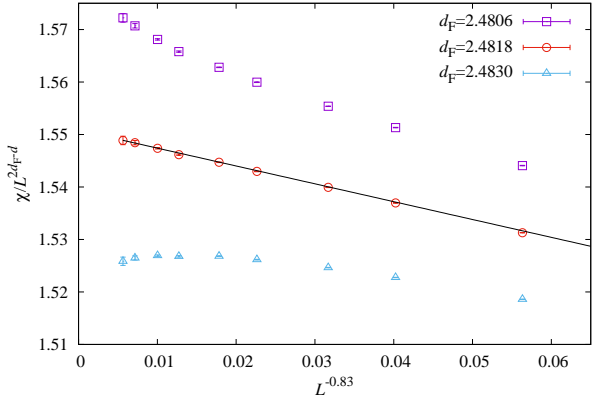}
\caption{Plots of $\chi$ vs $L^{-0.83}$ at $K_c$ for the 3D Ising model.}
\label{Fig:df-S2}
\end{figure}

From these fits, we get the estimate $\df=2.481\,8(4)$ (3D) and $\df = 1.875\,00(4)$ (2D), respectively.
In Fig.~\ref{Fig:df-S2}, we plot $\chi/ L^ {2\df-d}$
versus $L^{-0.83}$ using three different values of $\df$ for the 3D Ising model:
our estimate, as well as our estimate plus or minus three standard deviations.
As $L$ increases, the data with $\df=2.480\,6$ and $\df=2.483\,0$
bend upward and downward, respectively, while the data with $\df=2.481\,8$ are consistent with an asymptotically straight line.

\begin{table}
 \caption{Fits of $\chi$ for the 3D and 2D Ising models.}
 \scalebox{0.80}{
 \begin{tabular}[t]{|l|l|l|l|l|l|l|l|}
 \hline
                 	& $L_{\rm min}$ & $\chi^2$/DF & $\df$         & $a_0$         & $b_1$           & $y_1$       & $b_2$  \\
 \hline

 \multirow{5}{*}{3D}    &  8          & 5.6/8       & 2.481\,77(16)  & 1.551(2)  &$-0.30(2)$        & $-0.80(4)$  & $-0.86(8)$ \\
                	& 12          & 5.5/7       & 2.481\,7(3)    & 1.552(3)      &$-0.29(5)$        & $-0.78(8)$  & $-0.9(2)$  \\
                	&  8          & 6.2/9       & 2.481\,88(3)   & 1.549\,1(5)   &$-0.319(3)$       & $-0.83$     & $-0.79(2)$  \\
                 	& 12          & 5.9/8       & 2.481\,86(5)   & 1.549\,2(8)   &$-0.321(6)$       & $-0.83$     & $-0.77(5)$  \\
                	& 16          & 4.8/7       & 2.481\,82(7)   & 1.550(1)  &$-0.329(9)$       & $-0.83$     & $-0.7(1)$ \\
\hline
\multirow{5}{*}{2D} &  8          & 12.8/11     & 1.875\,00(2)   & 1.092\,1(2)   &$-0.14(2)$        & $-1.71(7)$  & ~~~-       \\
                	& 12          & 12.7/10     & 1.875\,00(2)   & 1.092\,0(2)   &$-0.16(6)$        & $-1.8(2)$   & ~~~-       \\
                 	& 16          & 10.4/10     & 1.875\,007(10) & 1.091\,9(1)   &$-0.30(2)$        & $-2$      & ~~~-       \\
                 	& 24          & 10.4/9      & 1.875\,007(13) & 1.091\,9(2)   &$-0.30(4)$        & $-2$      & ~~~-       \\
                 	& 32          &  7.0/8      & 1.875\,02(2)   & 1.091\,7(2)   &$-0.17(8)$        & $-2$      & ~~~-       \\
\hline
 \end{tabular}
}
\label{Tab:fit-S2}
\end{table}

\subsection{The second moment correlation length $\xi_{2nd}$}

At $K=K_c$, the ratio $\xi_{2nd}/L$
approaches a universal value $(\xi_{2nd}/L)_c$ in the thermodynamic limit $L \to \infty$.
It means that the second moment correlation length scales as $\xi_{2nd} \sim L$.
We fit the MC data of $\xi_{2nd}$ for the 3D and 2D Ising models
to Eq.(\ref{eq:A}) with $y_\scrA=1$ fixed and $a_0$ replaced
by $(\xi_{2nd}/L)_c$.
For the 3D Ising model, in the fit with $y_2=-2$ fixed and $y_1$ free,
we observe that the correction exponent $y_1 \approx -0.83$.
To reduce one fitting parameter, we further perform the fit
with $y_1=-0.83$ and $y_2=-2$ fixed.
For the 2D Ising model, if letting $b_2=0$ and $y_1$ free, we observe that the correction exponent $y_1 \approx -1.55$. 
To reduce one fitting parameter, we then try the fit with fixed $y_1=-1.55$. The fitting results are shown in Table~\ref{Tab:fit-CorLen}.

After comparing these fits, we determine the universal critical ratio as $(\xi_{2nd}/L)_c =0.6431(1)$ (3D)
and $0.905\,06(8)$ (2D), respectively. The current estimates for the 3D and 2D Ising models agree well with the previous
reported value $(\xi_{2nd}/L)_c (3\rm D)=0.643\,1(1)$~\cite{Hasenbusch2010a}, and the numerical integration result 
$(\xi_{2nd}/L)_c(2\rm D)=0.905\,048\,829\,2(4)$~\cite{Salas2000} using conformal field theory, respectively.

\begin{table}
\caption{Fits of $\xi_{2nd}$ for the 3D and 2D Ising models.}
 \scalebox{0.9}{
 \begin{tabular}[t]{|l|l|l|l|l|l|l|}
 \hline
					  &  $L_{\rm min}$      & $\chi^2$/DF      & $(\xi_{2nd}/L)_c$     & $b_1$                 & $y_1$        & $b_2$ \\
\hline
\multirow{5}{*}{3D}   &  12					&  10/8			   & 0.64321(9)			   &$-0.026(4)$			   & $-0.72(6)$   & $0.02(3)$ \\
 					  &  16					&  7.3/7		   & 0.64310(10)	       &$-0.037(10)$		   & $-0.83(9)$   & $0.11(7)$ \\
 					  &  16					&  7.4/8		   & 0.64310(3)	       	   &$-0.0367(8)$		   & $-0.83$  	  & $0.11(2)$ \\
 					  &  24					&  7.1/7		   & 0.64309(4)	       	   &$-0.036(1)$			   & $-0.83$  	  & $0.09(5)$ \\
 					  &  32					&  6.9/6		   & 0.64310(5)	           &$-0.037(2)$			   & $-0.83$   	  & $0.13(9)$ \\
 \hline
 \multirow{6}{*}{2D}  &   8                 &  6.4/12          & 0.905\,05(6)          &~~$0.46(2)$            & $-1.54(2)$   &  ~~~~-    \\
					  &  12                 &  6.3/11          & 0.905\,06(7)          &~~$0.48(4)$            & $-1.56(4)$   &  ~~~~-    \\
					  &  16                 &  6.2/10          & 0.905\,06(8)          &~~$0.47(7)$            & $-1.55(6)$   &  ~~~~-    \\
                      &   8                 &  6.5/13          & 0.905\,06(4)          &~~$0.468(2)$           & $-1.55$      &  ~~~~-    \\
                      &  12                 &  6.3/12          & 0.905\,05(4)          &~~$0.469(4)$           & $-1.55$      &  ~~~~-    \\
                      &  16                 &  6.2/11          & 0.905\,06(5)          &~~$0.468(7)$           & $-1.55$      &  ~~~~-    \\
 \hline
 \end{tabular}
  }
 \label{Tab:fit-CorLen}
 \end{table}


\begin{thebibliography}{38}
\expandafter\ifx\csname natexlab\endcsname\relax\def\natexlab#1{#1}\fi
\expandafter\ifx\csname bibnamefont\endcsname\relax
  \def\bibnamefont#1{#1}\fi
\expandafter\ifx\csname bibfnamefont\endcsname\relax
  \def\bibfnamefont#1{#1}\fi
\expandafter\ifx\csname citenamefont\endcsname\relax
  \def\citenamefont#1{#1}\fi
\expandafter\ifx\csname url\endcsname\relax
  \def\url#1{\texttt{#1}}\fi
\expandafter\ifx\csname urlprefix\endcsname\relax\def\urlprefix{URL }\fi
\providecommand{\bibinfo}[2]{#2}
\providecommand{\eprint}[2][]{\url{#2}}

\bibitem{Ising25} E. Ising, Z. Phys. {\bf 31}, 253 (1925).

\bibitem{Onsager44} L. Onsager, Phys. Rev. {\bf 64}, 117 (1944).

\bibitem{Baxter-book} R. J. Baxter, Exactly Solved Models in Statistical Mechanics (Academic Press, 1982).  

\bibitem[{\citenamefont{Ferrenberg et~al.}(2018)}]{FerrenbergXuLandau2018}
\bibinfo{author}{\bibfnamefont{A.~M.} \bibnamefont{Ferrenberg}},
\bibinfo{author}{\bibfnamefont{J.}~\bibnamefont{Xu}} \bibnamefont{and}
\bibinfo{author}{\bibfnamefont{D.~P.} \bibnamefont{Landau}},
\bibinfo{journal}{Phys. Rev. E} \textbf{\bibinfo{volume}{97}},
\bibinfo{pages}{043301} (\bibinfo{year}{2018}).
	

\bibitem{FK} 
P. W. Kasteleyn and C. M. Fortuin, J. Phys. Soc. Jpn.  {\bf 26} (Suppl.), 11 (1969); 
C. M. Fortuin and P. W. Kasteleyn Physica (Amsterdam) {\bf 57}, 536 (1972).

\bibitem[{\citenamefont{Swendsen and Wang}(1987)}]{SwendsenWang1987}
\bibinfo{author}{\bibfnamefont{R.~H.}~\bibnamefont{Swendsen}} \bibnamefont{and}
  \bibinfo{author}{\bibfnamefont{J.~S.} \bibnamefont{Wang}},
  \bibinfo{journal}{Phys. Rev. Lett.} \textbf{\bibinfo{volume}{58}},
  \bibinfo{pages}{86} (\bibinfo{year}{1987}).
  
 \bibitem[{\citenamefont{Edwards and Sokal}(1988)}]{EdwardsSokal1988}
\bibinfo{author}{\bibfnamefont{R.~G.}~\bibnamefont{Edwards}} \bibnamefont{and}
  \bibinfo{author}{\bibfnamefont{A.~S.} \bibnamefont{Sokal}},
  \bibinfo{journal}{Phys. Rev. D} \textbf{\bibinfo{volume}{38}},
  \bibinfo{pages}{2009} (\bibinfo{year}{1988}).

\bibitem[{\citenamefont{Langlands et~al.}(1992)}]{LanglandsPichetPouliotSaint-Aubin1992}
\bibinfo{author}{\bibfnamefont{R.~P.} \bibnamefont{Langlands}},
\bibinfo{author}{\bibfnamefont{C.} \bibnamefont{Pichet}},
\bibinfo{author}{\bibfnamefont{P.} \bibnamefont{Pouliot}} \bibnamefont{and}
\bibinfo{author}{\bibfnamefont{Y.} \bibnamefont{Saint-Aubin}},
\bibinfo{journal}{J. Stat. Phys.} \textbf{\bibinfo{volume}{67}},
\bibinfo{pages}{553} (\bibinfo{year}{1992}).

\bibitem[{\citenamefont{Pinson}(1994)}]{Pinson1994}
\bibinfo{author}{\bibfnamefont{H.~T.} \bibnamefont{Pinson}},
  \bibinfo{journal}{J. Stat. Phys.} \textbf{\bibinfo{volume}{75}},
  \bibinfo{pages}{1167} (\bibinfo{year}{1994}).

\bibitem[{\citenamefont{Arguin}(2002)}]{Arguin2002}
\bibinfo{author}{\bibfnamefont{L.~P.} \bibnamefont{Arguin}},
  \bibinfo{journal}{J. Stat. Phys.} \textbf{\bibinfo{volume}{109}},
  \bibinfo{pages}{301} (\bibinfo{year}{2002}).

\bibitem[{\citenamefont{Ziff et~al.}(1999)}]{ZiffLorenzKleban1999}
\bibinfo{author}{\bibfnamefont{R.~M.} \bibnamefont{Ziff}},
\bibinfo{author}{\bibfnamefont{C.~D.} \bibnamefont{Lorenz}} \bibnamefont{and}
\bibinfo{author}{\bibfnamefont{P.} \bibnamefont{Kleban}},
  \bibinfo{journal}{Physica A} \textbf{\bibinfo{volume}{266}},
  \bibinfo{pages}{17} (\bibinfo{year}{1999}).

\bibitem[{\citenamefont{Newman and Ziff}(2001)}]{NewmanZiff2001}
\bibinfo{author}{\bibfnamefont{M.~E.~J.}~\bibnamefont{Newman}} \bibnamefont{and}
  \bibinfo{author}{\bibfnamefont{R.~M.} \bibnamefont{Ziff}},
  \bibinfo{journal}{Phys. Rev. E} \textbf{\bibinfo{volume}{64}},
  \bibinfo{pages}{016706} (\bibinfo{year}{2001}).

\bibitem[{\citenamefont{Wang et~al.}(2013)}]{WangZhouZhangGaroniDeng2013}
\bibinfo{author}{\bibfnamefont{J.} \bibnamefont{Wang}},
\bibinfo{author}{\bibfnamefont{Z.} \bibnamefont{Zhou}},
\bibinfo{author}{\bibfnamefont{W.} \bibnamefont{Zhang}},
\bibinfo{author}{\bibfnamefont{T.~M.} \bibnamefont{Garoni}}, \bibnamefont{and}
\bibinfo{author}{\bibfnamefont{Y.}~\bibnamefont{Deng}},
  \bibinfo{journal}{Phys. Rev. E} \textbf{\bibinfo{volume}{87}},
  \bibinfo{pages}{052107} (\bibinfo{year}{2013}).

\bibitem[{\citenamefont{Wolff}(1989)}]{Wolff1989}
\bibinfo{author}{\bibfnamefont{U.} \bibnamefont{Wolff}},
  \bibinfo{journal}{Phys. Rev. Lett.} \textbf{\bibinfo{volume}{62}},
  \bibinfo{pages}{361} (\bibinfo{year}{1989}).

\bibitem{Xu14} X. Xu, J. F. Wang, Z. Zhou,  T. M. Garoni, Y. Deng, 
	Phys. Rev. E {\bf 89}, 012120 (2014).

\bibitem[{\citenamefont{Huang et~al.}(2018)}]{HuangHouWangZiffDeng2018}
\bibinfo{author}{\bibfnamefont{W.}~\bibnamefont{Huang}},
\bibinfo{author}{\bibfnamefont{P.~C.}~\bibnamefont{Hou}},
\bibinfo{author}{\bibfnamefont{J.~F.}~\bibnamefont{Wang}},
\bibinfo{author}{\bibfnamefont{R.~M.}~\bibnamefont{Ziff}} \bibnamefont{and}
\bibinfo{author}{\bibfnamefont{Y.~J.} \bibnamefont{Deng}},
\bibinfo{journal}{Phys. Rev. E} \textbf{\bibinfo{volume}{97}},
\bibinfo{pages}{022107} (\bibinfo{year}{2018}).

\bibitem[{\citenamefont{Munger and Novotny}(1991)}]{MungerNovotny1991}
\bibinfo{author}{\bibfnamefont{E.~P.}~\bibnamefont{M\"unger}} \bibnamefont{and}
  \bibinfo{author}{\bibfnamefont{M.~A.} \bibnamefont{Novotny}},
  \bibinfo{journal}{Phys. Rev. B} \textbf{\bibinfo{volume}{43}},
  \bibinfo{pages}{5773} (\bibinfo{year}{1991}).

\bibitem[{\citenamefont{Nienhuis}(1984)}]{Nienhuis1984}
\bibinfo{author}{\bibfnamefont{B.} \bibnamefont{Nienhuis}},
  \bibinfo{journal}{J. Stat. Phys.} \textbf{\bibinfo{volume}{34}},
  \bibinfo{pages}{731} (\bibinfo{year}{1984}).

\bibitem[{\citenamefont{di Francesco et~al.}(1987)}]{FrancescoSaleurZuber1987}
\bibinfo{author}{\bibfnamefont{P.} \bibnamefont{di Francesco}},
\bibinfo{author}{\bibfnamefont{H.} \bibnamefont{Saleur}} \bibnamefont{and}
\bibinfo{author}{\bibfnamefont{J.~B.} \bibnamefont{Zuber}},
  \bibinfo{journal}{J. Stat. Phys.} \textbf{\bibinfo{volume}{49}},
  \bibinfo{pages}{57} (\bibinfo{year}{1987}).

\bibitem{Cardy92} J. L. Cardy, J. Phys. A: Math. Gen. {\bf 25}, L201 (1992).



\bibitem[{\citenamefont{Hu and Deng}(2015)}]{HuDeng2015}
\bibinfo{author}{\bibfnamefont{H.}~\bibnamefont{Hu}} \bibnamefont{and}
  \bibinfo{author}{\bibfnamefont{Y.} \bibnamefont{Deng}},
  \bibinfo{journal}{Nucl. Phys. B.} \textbf{\bibinfo{volume}{898}},
  \bibinfo{pages}{157} (\bibinfo{year}{2015}).

\bibitem[{\citenamefont{Martins and Plascak}(2003)}]{MartinsPlascak2003}
\bibinfo{author}{\bibfnamefont{P.~H.~L.}~\bibnamefont{Martins}} \bibnamefont{and}
  \bibinfo{author}{\bibfnamefont{J.~A.} \bibnamefont{Plascak}},
  \bibinfo{journal}{Phys. Rev. E} \textbf{\bibinfo{volume}{67}},
  \bibinfo{pages}{046119} (\bibinfo{year}{2003}).

\bibitem[{\citenamefont{Deng and Bl\"ote}(2003)}]{DengBlote2003}
\bibinfo{author}{\bibfnamefont{Y.}~\bibnamefont{Deng}} \bibnamefont{and}
  \bibinfo{author}{\bibfnamefont{H.~W.~J.} \bibnamefont{Bl\"ote}},
  \bibinfo{journal}{Phys. Rev. E} \textbf{\bibinfo{volume}{68}},
  \bibinfo{pages}{036125} (\bibinfo{year}{2003}).

\bibitem[{\citenamefont{Hasenbusch}(2010)}]{Hasenbusch2010a}
\bibinfo{author}{\bibfnamefont{M.} \bibnamefont{Hasenbusch},},
  \bibinfo{journal}{Phys. Rev. B} \textbf{\bibinfo{volume}{82}},
  \bibinfo{pages}{174433} (\bibinfo{year}{2010}).


\bibitem{Yang52} C. N. Yang, Phys. Rev. {\bf 85}, 808 (1952).

\bibitem{Nienhuis87} B. Nienhuis, in Phase Transitions and Critical Phenomena, 
	edited by C. Domb and J. L. Lebowitz (Academic Press, New York, 1987), Vol. {\bf 11}.

\bibitem[{\citenamefont{Deng and Zhang}(2010)}]{DengZhang2010}
\bibinfo{author}{\bibfnamefont{Y.}~\bibnamefont{Deng}},
\bibinfo{author}{\bibfnamefont{W.} \bibnamefont{Zhang}},
\bibinfo{author}{\bibfnamefont{T.~M.} \bibnamefont{Garoni}},
\bibinfo{author}{\bibfnamefont{A.~D.} \bibnamefont{Sokal}} \bibnamefont{and}
\bibinfo{author}{\bibfnamefont{A.} \bibnamefont{Sportiello}},
\bibinfo{journal}{Phys. Rev. E} \textbf{\bibinfo{volume}{81}},
\bibinfo{pages}{020102} (\bibinfo{year}{2010}).

\bibitem[{\citenamefont{Deng and Bl\"ote}(2004)}]{DengBlote2004-1}
\bibinfo{author}{\bibfnamefont{Y.}~\bibnamefont{Deng}} \bibnamefont{and}
  \bibinfo{author}{\bibfnamefont{H.~W.~J.} \bibnamefont{Bl\"ote}},
  \bibinfo{journal}{Phys. Rev. E} \textbf{\bibinfo{volume}{70}},
  \bibinfo{pages}{046106} (\bibinfo{year}{2004}).

\bibitem[{\citenamefont{Deng et~al.}(2004)}]{DengBloteNienhuis2004-3}
\bibinfo{author}{\bibfnamefont{Y.}~\bibnamefont{Deng}},
  \bibinfo{author}{\bibfnamefont{H.~W.~J.} \bibnamefont{Bl\"ote}} \bibnamefont{and}
  \bibinfo{author}{\bibfnamefont{B.}~\bibnamefont{Nienhuis}},
  \bibinfo{journal}{Phys. Rev. E} \textbf{\bibinfo{volume}{69}},
  \bibinfo{pages}{026114} (\bibinfo{year}{2004}).

\bibitem{Eren16} E. M. Elci, M. Weigel, N. G. Fytas, Nucl. Phys. B {\bf 903}, 19 (2016).
\bibitem{Hu14} H. Hu, H. W. J. Bl\"ote, R. M. Ziff, Y. Deng, Phys. Rev. E {\bf 90}, 042106 (2014).

\bibitem[{\citenamefont{Tan et~al.}(2018)}]{Tan2018}
\bibinfo{author}{\bibfnamefont{X. J.}~\bibnamefont{Tan}},
  \bibinfo{author}{\bibfnamefont{R.} \bibnamefont{Couvreur}},
  \bibinfo{author}{\bibfnamefont{Y. J.} \bibnamefont{Deng}} \bibnamefont{and}
  \bibinfo{author}{\bibfnamefont{J. L.} \bibnamefont{Jacobsen}},
  \bibinfo{eprint}{arXiv:1809.06650} (\bibinfo{year}{2018}).

\bibitem[{\citenamefont{Couvreur and Jacobsen}(2017)}]{CouvreurJacobsen2017}
  \bibinfo{author}{\bibfnamefont{R.} \bibnamefont{Couvreur}},
  \bibinfo{author}{\bibfnamefont{J. L.} \bibnamefont{Jacobsen}} \bibnamefont{and}
  \bibinfo{author}{\bibfnamefont{R.} \bibnamefont{Vasseur}},
  \bibinfo{journal}{J. Phys. A: Math. Theor.} \textbf{\bibinfo{volume}{50}},
  \bibinfo{pages}{474001} (\bibinfo{year}{2017}).

\bibitem[{\citenamefont{EL-Showk et~al}(2012)}]{El-Showk2012}
  \bibinfo{author}{\bibfnamefont{S.} \bibnamefont{El-Showk}},
  \bibinfo{author}{\bibfnamefont{M. F.} \bibnamefont{Paulos}},
  \bibinfo{author}{\bibfnamefont{S.}~\bibnamefont{Rychkov}},
  \bibinfo{author}{\bibfnamefont{D.} \bibnamefont{Simmons-Duffin}} \bibnamefont{and}
  \bibinfo{author}{\bibfnamefont{A.} \bibnamefont{Vichi}},
  \bibinfo{journal}{Phys. Rev. D} \textbf{\bibinfo{volume}{86}},
  \bibinfo{pages}{025022} (\bibinfo{year}{2012}).

\bibitem[{\citenamefont{Poland et~al}(2018)}]{Poland2018}
\bibinfo{author}{\bibfnamefont{D.}~\bibnamefont{Poland}},
  \bibinfo{author}{\bibfnamefont{S.} \bibnamefont{Rychkov}} \bibnamefont{and}
  \bibinfo{author}{\bibfnamefont{A.} \bibnamefont{Vichi}},
  \bibinfo{eprint}{arXiv:1805.04405} (\bibinfo{year}{2018}).

\bibitem[{\citenamefont{Deng and Bl\"ote}(2005)}]{DengBlote2005}
\bibinfo{author}{\bibfnamefont{Y.}~\bibnamefont{Deng}} \bibnamefont{and}
  \bibinfo{author}{\bibfnamefont{H.~W.~J.} \bibnamefont{Bl\"ote}},
  \bibinfo{journal}{Phys. Rev. E} \textbf{\bibinfo{volume}{72}},
  \bibinfo{pages}{016126} (\bibinfo{year}{2005}).


 
\bibitem[{\citenamefont{Ferdinand and Fisher}(1969)}]{FerdinandFisher1969}
\bibinfo{author}{\bibfnamefont{A.~E.}~\bibnamefont{Ferdinand}} \bibnamefont{and}
  \bibinfo{author}{\bibfnamefont{M.~E.} \bibnamefont{Fisher}},
  \bibinfo{journal}{Phys. Rev.} \textbf{\bibinfo{volume}{185}},
  \bibinfo{pages}{832} (\bibinfo{year}{1969}).

\bibitem[{\citenamefont{Salas}(2001)}]{Salas2001}
\bibinfo{author}{\bibfnamefont{J.}~\bibnamefont{Salas}},
  \bibinfo{journal}{J. Phys. A: Math. Gen.} \textbf{\bibinfo{volume}{34}},
  \bibinfo{pages}{1311} (\bibinfo{year}{2001}).

\bibitem[{\citenamefont{Salas and Sokal}(2000)}]{Salas2000}
\bibinfo{author}{\bibfnamefont{J.}~\bibnamefont{Salas}} \bibnamefont{and}
  \bibinfo{author}{\bibfnamefont{A.~D.} \bibnamefont{Sokal}},
  \bibinfo{journal}{J. Statist. Phys.} \textbf{\bibinfo{volume}{98}},
  \bibinfo{pages}{551} (\bibinfo{year}{2000}).


\end{thebibliography}
\end{document}